\documentclass[aps,pre,superscriptaddress,nofootinbib,notitlepage,a4paper]{revtex4}

\usepackage{amsmath}
\usepackage{amsfonts}
\usepackage{amssymb}
\usepackage{xcolor}

\usepackage{graphicx}
\usepackage{float}
\usepackage{psfrag}
\usepackage[naturalnames]{hyperref}

\newcommand{\beq}{\begin{equation}} 
\newcommand{\eeq}{\end{equation}}
\newcommand{\bem}{\begin{multline}}
\newcommand{\bes}{\begin{split}}
\newcommand{\ees}{\end{split}} 
\newcommand{\bea}{\begin{align}}
\newcommand{\eea}{\end{align}}

\newcommand{\us}{\underline{\sigma}}

\newcommand{\hz}{\widehat{z}}

\newcommand{\uv}{\underline{v}}

\newcommand{\ad}{\alpha_{\rm d}}
\newcommand{\aalg}{\alpha_{\rm alg}}
\newcommand{\asat}{\alpha_{\rm sat}}
\newcommand{\e}{\epsilon}
\newcommand{\g}{\gamma}
\newcommand{\s}{\sigma}
\newcommand{\w}{\omega}
\newcommand{\dd}{{\rm d}}
\newcommand{\tpsi}{\widetilde{\psi}}
\newcommand{\tu}{\widetilde{u}}
\newcommand{\tw}{\widetilde{\omega}}
\newcommand{\tC}{\widetilde{C}}
\newcommand{\tH}{\widetilde{H}}

\newcommand{\halpha}{\widehat{\alpha}}
\newcommand{\he}{\widetilde{\epsilon}}
\newcommand{\heta}{\widehat{\eta}}
\newcommand{\hmu}{\widehat{\mu}}
\newcommand{\hnu}{\widehat{\nu}}
\newcommand{\hh}{\widehat{h}}

\newcommand{\hM}{\widehat{M}}
\newcommand{\hP}{\widehat{P}}
\newcommand{\hQ}{\widehat{Q}}

\newcommand{\hp}{\widehat{p}}
\newcommand{\hy}{\widehat{y}}
\newcommand{\ind}{\mathbb{I}}

\newcommand{\G}{{\cal G}}
\newcommand{\N}{{\cal N}}
\newcommand{\cP}{{\cal P}}
\newcommand{\eqd}{\overset{\rm d}{=}}

\newcommand{\di}{\partial i}
\newcommand{\da}{\partial a}
\newcommand{\dima}{\partial i \setminus a}
\newcommand{\dami}{\partial a \setminus i}
\newcommand{\mae}{\ \text{a.e.}}
\newcommand{\Po}{\text{Po}}
\newcommand{\ua}{u^{(1)}}
\newcommand{\ub}{u^{(2)}}
\newcommand{\uc}{u^{(3)}}
\newcommand{\tua}{\tu^{(1)}}
\newcommand{\tub}{\tu^{(2)}}
\newcommand{\tuc}{\tu^{(3)}}
\newcommand{\ea}{e^{(1)}}
\newcommand{\eb}{e^{(2)}}
\newcommand{\ec}{e^{(3)}}
\newcommand{\za}{z^{(1)}}
\newcommand{\zb}{z^{(2)}}
\newcommand{\zc}{z^{(3)}}
\newcommand{\Vbar}{\overline{V}}

\begin{document}

\bibliographystyle{myunsrt}

\title{Biased measures for random Constraint Satisfaction Problems:\\larger interaction range and asymptotic expansion}

\author{Louise Budzynski}
\affiliation{Laboratoire de physique de l'Ecole normale sup\'erieure, ENS, Universit\'e PSL, CNRS, Sorbonne Universit\'e, Universit\'e de Paris, F-75005 Paris, France} 

\author{Guilhem Semerjian}
\affiliation{Laboratoire de physique de l'Ecole normale sup\'erieure, ENS, Universit\'e PSL, CNRS, Sorbonne Universit\'e, Universit\'e de Paris, F-75005 Paris, France} 

\begin{abstract}
We investigate the clustering transition undergone by an exemplary random constraint satisfaction problem, the bicoloring of $k$-uniform random hypergraphs, when its solutions are weighted non-uniformly, with a soft interaction between variables belonging to distinct hyperedges. We show that the threshold $\alpha_{\rm d}(k)$ for the transition can be further increased with respect to a restricted interaction within the hyperedges, and perform an asymptotic expansion of $\alpha_{\rm d}(k)$ in the large $k$ limit. We find that $\alpha_{\rm d}(k) = \frac{2^{k-1}}{k}(\ln k + \ln \ln k + \gamma_{\rm d} + o(1))$, where the constant $\gamma_{\rm d}$ is strictly larger than for the uniform measure over solutions.

\end{abstract}

\maketitle

\tableofcontents

\section{Introduction}

In a Constraint Satisfaction Problem (CSP) $N$ discrete valued variables are subject to $M$ constraints. Each of the constraints enforces some requirement on a subset of the variables, a solution of the CSP is thus an assignement of the variables that satisfies simultaneously all the constraints. Famous examples of CSPs are the $k$-satisfiability problem and the graph $q$-coloring one; we will focus in this paper on another related problem, the bicoloring of $k$-hypergraphs. In this problem the variables are boolean and lie on the vertices of an hypergraph, with hyperedges linking subsets of $k$ vertices (instead of two for a graph). The constraint associated to each hyperedge is that both values (or colors) are present among its $k$ adjacent vertices, forbidding locally monochromatic configurations.

CSPs have been studied from various perspectives. Computational complexity theory~\cite{GareyJohnson79,Papadimitriou94} classifies them according to their worst-case difficulty, characterized by the existence or not of an efficient (running in a time polynomial in the size $N$) algorithm able to solve (i.e. decide whether they admit a solution) all their possible instances. Another approach, in which this work takes place, focuses on the typical difficulty of CSPs, where typical is defined with respect to a random ensemble of instances (see e.g.~\cite{MonassonZecchina99b,BiroliMonasson00,MezardParisi02,MertensMezard06,krzakala2007gibbs,AchlioptasRicci06,AchlioptasCoja-Oghlan08,molloy_col_freezing,ding2014proof}). The most commomly studied random ensemble is obtained by drawing the $M$ constraints uniformly at random, i.e. by constructing a $G(N,M)$ Erd\H os-R\'enyi random $k$-hypergraph. In this paper we will focus on a slightly different ensemble, the $(k,l+1)$-regular one, where the probability is uniform on the set of hypergraphs for which each vertex belongs to $l+1$ hyperedges. A striking property of random CSPs is the occurence of phase transitions, or threshold phenomena, in the large size (thermodynamic) limit $N,M\to\infty$ with a fixed ratio $\alpha=M/N$ (in the regular ensemble $\alpha=(l+1)/k$), the density of constraints per variable. When the control parameter $\alpha$ is varied one observes several phase transitions, at which the probability of some properties jumps abruptly from 1 to 0 in the thermodynamic limit. In particular the satisfiability threshold $\asat$ separates an underconstrained regime $\alpha<\asat$ where typical instances are satisfiable, from an overconstrained regime $\alpha>\asat$ where they typically do not admit any solution. The existence of such a transition has been proven (in a slightly weaker sense) in~\cite{Friedgut99}, as well as lower~\cite{transition_lb,Achltcs} and upperbounds~\cite{transition_ub} on the threshold $\asat$, that become tighter and tighter as $k$ grows~\cite{AchlioptasMoore02,AchlioptasPeres04}. Statistical mechanics methods, adapted from the study of spin-glasses~\cite{MezardParisi87b,MezardMontanari07}, provided predictions for the value of $\asat(k)$ for several random CSPs~\cite{MezardParisi02,MertensMezard06,KrzakalaPagnani04}, the correctness of this method was later proven rigorously for large but finite $k$~\cite{DiSlSu13_naeksat,ding2014proof}. There appears to be a large universality among the various random CSPs that have been studied, in particular in the large $k$ limit; for the sake of readability the quantitative statements below are given with the scale of $\alpha$ corresponding to the bicoloring problem, even when quoting papers that derived this result for another CSP.

Several other phase transitions occur in the satisfiable phase $\alpha < \asat$~\cite{krzakala2007gibbs}. In this paper we focus on the clustering phase transition $\ad$, which is also known as the dynamic or reconstruction transition. This transition can be defined in several ways. Below $\ad$ the set of solutions of most instances is rather well-connected, any typical solution can be reached from another one through a path constituted of nearby solutions. Above $\ad$ the solution set splits into an exponential number of isolated subsets of solutions, called clusters, which are internally well-connected but separated one from each other by regions without solutions: this is called the clustering phenomenon. This transition also manifests itself with the appearance of a specific form of long range correlations between variables, known as point-to-set correlations, in the probability law defined as the uniform measure over the set of solutions. These correlations imply the solvability of an information-theoretic problem called tree reconstruction~\cite{MoPe03,MezardMontanari06}, and forbid the rapid equilibration of the stochastic processes that respect the detailed balance condition~\cite{MontanariSemerjian06b},  which justifies the alternative ``dynamic'' name of the clustering transition. In the cavity method~\cite{MezardParisi01} treatment of the random CSPs $\ad$ can also be defined as the appearance of a non trivial solution of the one step of Replica Symmetry Breaking (1RSB) equation with Parisi breaking parameter $m=1$, see in particular~\cite{MezardMontanari06} for the connection between this formalism and the reconstruction problem. In the large $k$ limit the dynamic transition happens at a much smaller constraint density than the satisfiablity one, the asymptotic expansion of these two thresholds being $\ad(k)\sim 2^{k-1}\ln k / k$ and $\asat(k) \sim 2^{k-1}\ln 2$.

An important open problem in the field of random CSPs concerns the behavior of algorithms in the satisfiable regime, where the goal is to find a solution, as typical instances admit such configurations. In particular one would like to determine the algorithmic threshold $\aalg(k)$ above which no algorithm is able to find a solution in polynomial time with high probability (assuming P$\neq$NP). For small values of $k$ it is possible to design algorithms (see \cite{SelmanKautz94,MezardParisi02,ArdeliusAurell06,AlavaArdelius07,MaPaRi15}) that are found through numerical simulations to be efficient at densities very close to the satisfiability threshold. Unfortunately these algorithms cannot be studied numerically in the large $k$ limit and one has to resort to analytical studies in this case, which can only be performed on relatively simple heuristics. The best result in this direction is the one of~\cite{Amin_algo}, which provides an algorithm that provably works in polynomial time up to constraint densities of the order of $2^{k-1}\ln k / k$, i.e. the same asymptotic scaling as $\ad(k)$. This leaves a wide range of $\alpha$ where typical instances have a non-empty set of solutions, but no known algorithm is able to find them efficiently (and where some families of algorithms have been proven to fail~\cite{GaSu14,CoHaHe17,Hetterich}).

One could hope that this algorithmic question, and in particular the value of $\aalg(k)$, could be enlightened by the accumulated knowledge on the several structural phase transitions that occur in the satisfiable phase. The connection between these two aspects is however very delicate because algorithms are intrisically out-of-equilibrium processes, either because their mere definition violates the detailed balance condition, or because they fall out of equilibrium during their evolution on a time scale that is shorter than their relaxation time. In both cases there are no fundamental principles to connect their dynamics with the static properties of the solution set. Even if one cannot understand precisely $\aalg(k)$ in terms of a structural phase transition one can reasonably state that the dynamic transition is a lower bound to the algorithmic one, $\ad(k) \le \aalg(k)$. Indeed for $\alpha \le \alpha_{\rm d}$ simulated annealing~\cite{KirkpatrickGelatt83} should be able to reach thermalization in polynomial time down to arbitrarily small temperatures, and hence sample uniformly the solution set. For $\alpha$ slightly larger than $\alpha_{\rm d}$ one expects simulated annealing to fall out-of-equilibrium on polynomial timescales but in many cases it should still be able to find (non-uniformly) solutions, hence the bound $\ad(k) \le \aalg(k)$ is not tight in general.

The study of the structural phase transitions in the satisfiable regime, and in particular the definition of $\ad$ in terms of long-range correlations, relies on the characterization of a specific probability law on the space of configurations, namely the uniform measure over solutions. This paper belongs to a series of works studying a probability measure over the set of solutions that is biased instead of uniform, i.e. that weights differently the various solutions of the CSP instance. This idea has been used in several articles, see in particular~\cite{BrDaSeZd16,BaInLuSaZe15_long,BaBo16,MaSeSeZa18,BuRiSe19,ZhZh20}, with slightly different perspectives and results (for instance in~\cite{BaInLuSaZe15_long,BaBo16} solutions are weighted according to the number of other solutions in their neighborhood, in~\cite{BrDaSeZd16} according to their number of frozen variables taking the same value in the whole cluster, while in~\cite{MaSeSeZa18} the solutions are non-overlapping positions of hard spheres, biased through an additional pairwise soft interaction between them). In~\cite{BuRiSe19} we have studied a simple implementation of the bias in the measure on the set of solutions of an hypergraph bicoloring instance, where the interactions induced by the bias can be factorized over the bicoloring constraints, and studied systematically the modifications of the clustering threshold $\ad$ induced by the non-uniformity between solutions. We showed, for $k$ between 4 and 6, that with well-chosen parameters such a bias allows to increase $\ad$, and to improve the performances of simulated annealing, in agreement with the discussion above. However in~\cite{BuRiSe19} we left essentially open the question of the increase of $\ad$ this bias could achieve in the large $k$ limit, and hence whether such a strategy could reduce the algorithmic gap in this limit.

As a matter of fact the large $k$ behavior of $\ad$ is a rather involved asymptotic expansion, even for the uniform measure, and until recently only relatively loose bounds on the asymptotic behavior of $\ad$ were known~\cite{Sly08,MoReTe11_recclus,SlyZhang16}. We considered this specific problem in~\cite{BuSe19} and found that the clustering threshold occurs on the scale $\alpha \sim2^{k-1}(\ln k+\ln\ln k +\g )/k$ with $\g$ constant, and more precisely that for the uniform measure $\g_{\rm d,u}\approx 0.871$, which falls into the range allowed by the previous bounds~\cite{Sly08,MoReTe11_recclus,SlyZhang16}.

In this paper we build upon our previous works~\cite{BuRiSe19,BuSe19} and generalize them to obtain two main new results. We first introduce a more generic way of weighting the different solutions of an instance of the hypergraph bicoloring problem, that extends the one presented in~\cite{BuRiSe19} and incorporates interactions between variables belonging to different hyperedges, and shows that for finite $k$ it allows a further increase of the dynamic threshold $\ad$. Moreover we adapt the large $k$ expansion of~\cite{BuSe19} to this biased measure and manage to assess the asymptotic effect of the bias on $\ad$; we find that the factorized bias of~\cite{BuRiSe19} cannot improve on the constant $\g_{\rm d}$ in the asymptotic expansion with respect to $\g_{\rm d,u}$, while the bias with larger interaction range we introduce here allows to increase its value up to $\g_{\rm d} \approx 0.977$. This is arguably a modest improvement, bearing on the third order of the asymptotic expansion of $\ad$, nevertheless it opens the possibility to study further generalizations of the bias and to bring some light on the nature of the algorithmic gap between $\aalg$ and $\asat$.

The rest of the paper is organized as follows. In Section~\ref{sec:definitions} we define more precisely the bicoloring problem and the biased measure we introduce on its set of solutions. Its treatment with the simplest version of the cavity method is presented in Section~\ref{sec_RS}, while Sec.~\ref{sec_dynamic} refines this description to incorporate the clustering phenomenon and presents the equations that allow to compute the dynamic threshold. In section~\ref{sec:finitek_results} we display some numerical results for finite values of $k$ and compare them to the simpler biasing strategy of~\cite{BuRiSe19}. The analytical expansion of the dynamic transition threshold in the large $k$ limit is the subject of section~\ref{sec_largek}, followed by some conclusions and perspectives for future work in Sec.~\ref{sec_conclusion}. More technical details of the computations are deferred to the Appendices~\ref{app_uniqueness_RS} and~\ref{app_inequalities}.

\section{Definitions}
\label{sec:definitions}

\subsection{Biased measures with interactions at distance 1}
\label{sec_def_biased}

An instance of the bicoloring problem is specified by a $k$-uniform hypergraph $G=(V,E)$, where $V=\{1,\dots,N\}$ is a set of $N$ vertices and $E$ a set of $M$ hyperedges, each hyperedge $a \in E$ linking a subset denoted $\partial a$ of $k$ vertices (we shall denote similarly $\partial i$ the set of hyperedges in which a vertex $i$ appears); a graphical representation for a small example can be found in the left panel of Fig.~\ref{fig:factor_graph_loops}. Binary variables, encoded as Ising spins $\s_i \in \{-1,+1\}$, are placed on the vertices of $G$, their global configuration being denoted $\us=(\s_1,\dots,\s_N) \in \{-1,+1\}^N$; we shall write $\us_S=\{\s_i \, : \, i \in S \}$ for the configuration of the variables in an arbitrary subset $S \subset V$ of the vertices. A solution of the bicoloring problem (also called proper bicoloring of $G$) is, by definition, an assignment $\us$ of the variables such that no hyperedge $a \in E$ is monochromatic, in other words such that for each $a \in E$ there is at least one neighboring vertex $i \in \partial a$ with $\s_i=+1$, and at least one $i \in \partial a$ with $\s_i=-1$. Assuming that $G$ admits proper bicolorings (i.e. that the bicoloring problem on $G$ is satisfiable), we introduce the uniform measure over the set of solutions as
\beq
\label{eq:measure_uniform}
\rho_{\rm u}(\us) = \frac{1}{Z} \prod_{a \in E} \w(\us_{\da}) \ ,
\eeq
where the subscript $\rm u$ stands for ``uniform'', the normalization constant (partition function) $Z$ counts the number of solutions, and $\w(\s_1\dots\s_k)$ is the indicator function of the event ``$\s_1\dots\s_k$ are not all equal'':
\beq
\w(\s_1,\dots\s_k) = \begin{cases}
0 & \text{if} \quad \s_1=\dots=\s_k \ , \\
1 & \text{otherwise}  \ .
\end{cases}
\eeq

Our goal now is to define a measure $\rho$ which, as $\rho_{\rm u}$, has for support the set of solutions ($\rho(\us)>0$ if and only if $\us$ is a proper bicoloring), but that can give different weights to different solutions. There are obviously an infinite number of $\rho$ that fulfills this condition, we shall progressively narrow down these possibilities to arrive at the form of $\rho$ studied in the rest of the paper. For simplicity we restrict ourselves from now on to regular hypergraphs, where every vertex has the same degree $|\partial i|=l+1$.

We will first impose a locality requirement for $\rho$: for this biased measure to be tractable in Monte Carlo simulations or in analytical computations the additional interactions between variables induced by the non-uniformity should be local, with respect to the notion of distance induced by the hypergraph $G$. Considering the shortest non-trivial interaction range that allows the coupling of variables from different hyperedges yields the form
\beq
 \rho(\us) = \frac{1}{Z} \prod_{a \in E} \w(\us_{\da}) \prod_{i=1}^N \varphi\left (\s_i,\{\us_{\dami}\}_{a\in\di}\right) \ ,
\eeq
where the biasing function $\varphi > 0$ couples the $i$-th variable with its $(l+1)(k-1)$ neighbors at distance 1; as $G$ is regular and uniform we use the same function $\varphi$ on all the vertices.

There is still a vast freedom in the choice of $\varphi$; we further restrict it by imposing its invariance under the spin-flip symmetry $\us \to - \us$ (that is fulfilled by the set of solutions), and under the permutations of the $l+1$ hyperedges around $i$, and of the $k-1$ neighboring variables inside each of these hyperedges. This amounts to take
\beq
\varphi\left (\s_i,\{\us_{\dami}\}_{a\in\di}\right) = \widehat{\varphi}(\{m_{a \to i}\}_{a\in\di}) \qquad
\text{with} \ \ \ m_{a \to i} = \sum_{j\in\dami}\frac{1+\s_i\s_j}{2} \ ,
\eeq
and $\widehat{\varphi}$ invariant under the permutations of its $l+1$ arguments.

$m_{a \to i}$ counts the number of variables in $\dami$ that are of the same color as $\s_i$; we shall finally discard part of the information contained in $m_{a \to i}$, and only distinguish between the cases $m_{a \to i}=0$ and $m_{a \to i}>0$. This is indeed a relevant information about the solution $\us$: in the former case $\s_i$ is the only variable of its color in the $a$-th hyperedge, hence $\s_i$ cannot be flipped without violating the $a$-th monochromatic constraint, one says that $a$ forces $i$ in such a situation. On the contrary when $m_{a \to i} > 0$ the variable $\s_i$ is not forced to its value by the $a$-th hyperedge. With this simplification, and because of the invariance by permutation of the arguments of $\widehat{\varphi}$, the weight on the variable $i$ becomes a function of the number of constraints forcing it. As $\us$ is a solution the event $m_{a \to i}=0$ is equivalent to ``the variables in $\us_{\dami}$ are all equal (a.e.)'', the biased measure becomes thus
\beq
\rho(\us) = \frac{1}{Z} \prod_{a \in E} \w(\us_{\da}) \prod_{i=1}^N \psi\left(\sum_{a\in\di} \ind[\us_{\dami} \mae] \right)\ ,
\label{eq:measure_withbiasdistance1}
\eeq
where here and in the following $\ind[A]$ is the indicator function of the event $A$, and $\psi(p)>0$ is the weight attributed to a variable contained in $p\in\{0,\dots,l+1\}$ forcing hyperedges.

The following study will be devoted to the properties of the measure (\ref{eq:measure_withbiasdistance1}); despite the several restricting hypotheses we have made to reach this form of $\rho$ there are still $l+1$ free parameters in $\psi$ (its argument can take $l+2$ values, but a global multiplicative constant gets absorbed in the normalization $Z$). Part of our computations will be made for an arbitrary $\psi$ but in some places, and in particular in the large $k$ analysis, we will use the specific form:
\beq
\psi(0)=1 \ , \quad \psi(p)=b (1-\epsilon)^p \quad \text{for} \quad 1\leq p\leq l+1 \ ,
\label{eq_bias_beps}
\eeq
which contains the two parameters $b>0$ and $\epsilon <1$. This form actually encompasses two cases previously investigated in the literature:
\begin{itemize}
\item when $\epsilon=0$, in such a way that
\beq
\label{eq:bias_BrDaSeZd16}
\psi(p)= \begin{cases}
1 & \text{if } p=0 \ , \\
b & \text{if } p>0 \ ,
\end{cases}
\eeq
one recovers a measure studied in~\cite{BrDaSeZd16}. Indeed the weight of a solution $\us$ is then $b$ raised to a power equal to the number of variables that are forced by at least one constraint, i.e. those that are not ``whitened'' after $T=1$ step of a coarsening algorithm used in~\cite{BraunsteinMezard05,BraunsteinMezard02,Parisi02b,BraunsteinZecchina04,ManevaMossel05,AchlioptasRicci06,DiSlSu13_naeksat}, whose large deviations on atypical solutions were studied in~\cite{BrDaSeZd16} for arbitrary values of $T$ (with an unfortunate conflict of notation the parameter called $b$ here is denoted $e^\epsilon$ in~\cite{BrDaSeZd16}).

\item when $b=1$ one has 
\beq
\label{eq:bias_factorized}
\psi(p)=(1-\epsilon)^p \ ,	
\eeq
the weight of a solution $\us$ is thus $(1-\epsilon)$ raised to the number of forcing constraints. Indeed in a solution every constraint $a$ forces at most one of its variable $i \in \da$, when $\s_i$ is the unique representant of its color in $\us_{\da}$, hence there is no double counting of forcing constraints in the product over variables in (\ref{eq:measure_withbiasdistance1}). This case was investigated in~\cite{BuRiSe19}, and is somehow simpler thanks to the factorization of the biasing function along the hyperedges of $G$.
\end{itemize}
Obviously the uniform measure (\ref{eq:measure_uniform}) is recovered when $b=1$ and $\epsilon=0$.

\subsection{Factor graph and auxiliary variables}

\begin{figure}
\begin{center}
\includegraphics[width=14cm]{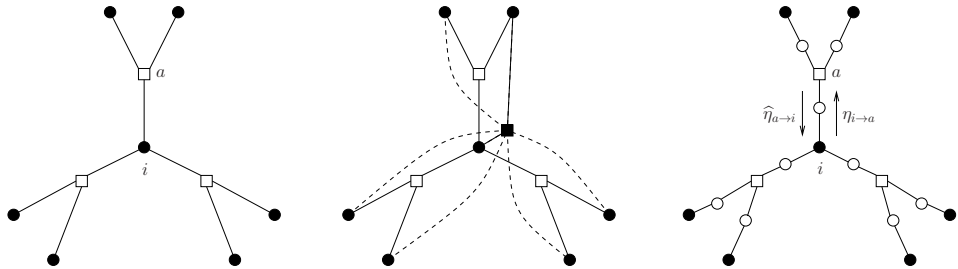}
\caption{Left panel: an example of an hypergraph $G$ with $N=7$ vertices represented by black circles, and $M=3$ hyperedges linking $k=3$ vertices, drawn as white squares. Center panel: the introduction of an interaction, represented as a black square, between all the vertices at distance 1 from the central vertex $i$, generates short loops even if $G$ is a tree. Right panel: the factor graph representation of the probability measure (\ref{eq:measure_on_ux}), the white circles stand for the variable nodes $v_{(i,a)}$, the black circles (resp. white squares) are the interaction factors $\tpsi$ (resp. $\tw$). The messages $\eta_{i \to a}$ and $\heta_{a \to i}$ obey the Belief Propagation equations (\ref{eq:BP_f_g}).}
 \label{fig:factor_graph_loops}
\end{center}
\end{figure}

We will study the properties of the measure (\ref{eq:measure_withbiasdistance1}) when $G$ is drawn uniformly at random among all $k$-uniform, $l+1$-regular hypergraphs, in the thermodynamic limit $N,M \to \infty$ with $Mk=N(l+1)$; in this limit such hypergraphs converge locally to hypertrees (they contain a bounded number of finite length cycles), which allows the use of the Belief Propagation algorithm~\cite{Pearl88,KschischangFrey01,YedidiaFreeman03} and of the cavity method analysis~\cite{MezardParisi01,MezardMontanari07}, that are based on this locally tree-like structure. However the probability law (\ref{eq:measure_withbiasdistance1}), interpreted as a graphical model with variables $\s_i$, contain short loops even if $G$ is a tree, because the biasing term $\psi$ couples all the variables at distance 1 from each vertex (see the middle panel of Fig.~\ref{fig:factor_graph_loops}). This prevents a direct application of the cavity method, and requires a preliminary step in order to get rid of these short loops. To achieve this we introduce some auxiliary, redundant variables in the following way: for each edge $(i,a)$ between a vertex $i$ and one of its incident hyperedge $a \in \di$ we introduce two variables, $w_{a \to i}\in \{0,1\}$ and $\s_i^a \in \{-1,+1\} $, which are deterministic functions of the original configuration $\us$, according to $w_{a \to i} = \ind[\us_{\dami} \mae]$ and $\s_i^a=\s_i$. We will call $v_{(i,a)} = (\s_i^a,w_{a \to i})$ the value of these two auxiliary variables, and $\uv=\{v_{(i,a)}\}_{i \in V,a \in \di}$ their global configuration. Consider now the following probability law for $\uv$:
\beq
\label{eq:measure_on_ux}
\rho(\uv)=\frac{1}{Z}\prod_{i=1}^N\tpsi(\{v_{(i,a)}\}_{a\in\di})\prod_{a \in E}\tw(\{v_{(i,a)}\}_{i\in\da}) \ ,
\eeq
where
\begin{align}
& \tpsi(\s_1,w_1,\dots,\s_{l+1},w_{l+1}) = \psi\left(\sum_{i=1}^{l+1} w_i\right)\ind[\s_1=\dots=\s_{l+1}] \ , 
\label{eq_tpsi} \\
& \tw(\s_1,w_1,\dots,\s_k,w_k) = \w(\s_1,\dots,\s_k)\prod_{i=1}^k\ind\left[w_i=\ind[\{\s_j\}_{j\neq i} \mae ]\right]  \ . \label{eq_tw}
\end{align}
One realizes easily that for a configuration $\uv$ in the support of (\ref{eq:measure_on_ux}) $\s_i^a$ is independent of $a$, and that the marginal law of $\us$ is nothing but (\ref{eq:measure_withbiasdistance1}). The partition function $Z$ is the same in the two expressions (\ref{eq:measure_withbiasdistance1}) and (\ref{eq:measure_on_ux}), and in the support of $\rho(\uv)$ the variables $w_{a \to i}$ are the deterministic functions of $\us$ defined above. This equivalent rewriting with redundant variables has an important advantage: as shown in the right panel of Fig.~\ref{fig:factor_graph_loops} the graphical model corresponding to (\ref{eq:measure_on_ux}), with variables $v$ on the edges $(i,a)$ of $G$ and interaction nodes both on the original hyperedges ($\tw$) and on the original vertices ($\tpsi$), respects the topology of $G$, and is thus (locally) a tree if $G$ is. 

\section{The replica symmetric cavity formalism}
\label{sec_RS}
\subsection{Belief Propagation Equations}
\label{subsec:BP_equations}

We will study the typical properties of the probability law (\ref{eq:measure_on_ux}) for large random hypergraphs with the cavity method~\cite{MezardParisi01,MezardMontanari07}. We first briefly recall the main ideas that underly it: if $G$ were a tree then all the marginals of (\ref{eq:measure_on_ux}), as well as the normalization constant $Z$, could be efficiently computed by recursion, breaking the tree into independent subtrees and combining the results of the computations in the smaller substructures. This procedure takes the form of local recursion relations between ``messages'' passed from one variable node to its adjacent interaction nodes, and vice-versa, where these messages are probability laws for the variables in amputated factor graphs where some nodes have been removed. These local recursions are exact if the factor graph is a tree, they can nevertheless be used even if it has some cycles; in that case the corresponding algorithm, called Belief Propagation (BP)~\cite{Pearl88,KschischangFrey01,YedidiaFreeman03}, is a priori only approximate, with no convergence guarantee. Sparse random graph models being locally tree-like, BP is a good candidate to describe asymptotically their behavior. This is indeed the case within an hypothesis of correlation decay, called Replica Symmetry (RS), which implies that the long cycles of typical graphs do not spoil the locally tree-like features captured by BP. This RS hypothesis breaks down for frustrated enough models, in particular for constraint satisfaction problems at high enough densities, as we shall discuss in more details in the next Section.

For now let us state the form of the BP equations and of the RS cavity predictions for the model at hand. The BP messages are $\eta_{i\to a}$ and $\heta_{a\to i}$, the marginal laws of the variable $v_{(i,a)}$ that is placed on the edge $(i,a)$ of $G$, in graphs where one has removed the interactions $a$ and $i$, respectively. These definitions are illustrated in the right panel of figure~\ref{fig:factor_graph_loops}. Note that in a literal application of the BP algorithm one would have introduced messages from every variable node to every interaction node, for instance $\eta_{i \to (i,a)}$ and $\eta_{(i,a) \to a}$; as the variable nodes are of degree two these two messages are actually equal, we denoted their common value $\eta_{i \to a}$ to lighten the notations. The BP equations between these messages are of the form
\begin{align}
\label{eq:BP_f_g}
\eta_{i\to a} = f(\{\heta_{b\to i}\}_{b\in\dima}) \ , \quad \heta_{a\to i} = g(\{\eta_{j\to a}\}_{j\in\dami}) \ ,
\end{align}
where the functions $f$ and $g$ derive from the interaction nodes $\tpsi$ and $\tw$ stated in (\ref{eq_tpsi},\ref{eq_tw}). The relation $\eta=f(\heta_1,\dots,\heta_l)$ is thus found to mean
\begin{equation}
\label{eq:BP_eta}
\eta(\s,w) = \frac{1}{z}\sum_{w_1,\dots,w_l}\psi\left(w + \sum_{i=1}^l w_i \right)\prod_{i=1}^l \heta_i(\s,w_i) \ ,
\end{equation}
where $z$ is a normalization constant. Similarly $\heta=g(\eta_1,\dots,\eta_{k-1})$ stands for
\begin{align}
\label{eq:BP_heta}
\heta(\s,w) = &\frac{1}{\hz}\sum_{\substack{\s_1,\dots,\s_{k-1} \\ w_1,\dots,w_{k-1} }}\w\left(\s,\s_1,\dots,\s_{k-1} \right)\ind\left[w=\ind[ \s_1,\dots,\s_{k-1} \mae]\right] \\
& \prod_{i=1}^{k-1}\eta_i(\s_i,w_i)\ind\left[w_i=\ind[\s,\s_1,\dots , \s_{i-1} , \s_{i+1},\dots,\s_{k-1}  \mae ]\right] \ , \nonumber
\end{align}
with $\hz$ a normalization constant. More explicitly one has
\begin{align}
\heta(\s,1) &= \frac{1}{\hz} \prod_{i=1}^{k-1}\eta_i(-\s,0) \ , \\
\heta(\s,0) &= \frac{1}{\hz} \left[\sum_{i=1}^{k-1}\eta_i(-\s,1)\prod_{j\neq i}\eta_j(\s,0) + \sum_{\substack{I\subset\{1, \dots , k-1\}  \\ 2\leq|I|\leq k-2}}\prod_{i\in I}\eta_i(-\s,0) \prod_{i\not\in I}\eta_i(\s,0)\right] \ ,
\end{align}
as there is at most one variable which is the unique representant of its color in a set of $k \ge 3$ binary variables that is not monochromatic.

\subsection{The replica symmetric solution and thermodynamics}
\label{sec_RS_solution}

In a $k$-uniform $l+1$-regular hypergraph the local neighborhood of every vertex is the same, it is thus natural to look for a translationally invariant solution of the BP equations. Moreover the probability measure we are studying is invariant under the spin-flip symmetry $\us \to -\us$, we can thus further restrict ourselves to a solution of the BP equation that respects this invariance. This amounts to take $\eta_{i\to a}(\s,w)=\eta_*(w)$, $\heta_{a\to i}(\s,w)=\heta_*(w)$ for all edges $(i,a)$. Plugging this form into (\ref{eq:BP_f_g}) yields the equations satisfied by $\eta_*$ and $\heta_*$:
\begin{align}
\label{BP_invariance_spin_transl_1}
\eta_*(w) &= \frac{1}{z}\sum_{p=0}^l \binom{l}{p} \psi(p+w) \heta_*(0)^{l-p}\heta_*(1)^p \ , \\
\label{BP_invariance_spin_transl_2}
\heta_*(1) &= \frac{1}{\hz} \eta_*(0)^{k-1} \ , \\
\label{BP_invariance_spin_transl_3}
\heta_*(0) &= \frac{1}{\hz}\left[(k-1)\eta_*(1)\eta_*(0)^{k-2} + (2^{k-1}-k-1)\eta_*(0)^{k-1}\right] \ .
\end{align}
Introducing the ratio of probabilities 
\beq
y=\frac{\eta_*(0)}{\eta_*(1)} \ ,  \qquad \hy=\frac{\heta_*(0)}{\heta_*(1)} \ ,
\label{eq_def_y}
\eeq
one can get rid of the normalization factors $z$ and $\hz$ and rewrite (\ref{BP_invariance_spin_transl_1},\ref{BP_invariance_spin_transl_2},\ref{BP_invariance_spin_transl_3}) more simply
\beq
y = \frac{\underset{p=0}{\overset{l}{\sum}}  \binom{l}{p} \psi(p) \hy^{-p}}{\underset{p=0}{\overset{l}{\sum}}  \binom{l}{p} \psi(p+1) \hy^{-p}} \ , \qquad \hy = 2^{k-1} -k-1 + \frac{k-1}{y} \ .
\label{BP_invariance_spin_transl_with_yhy}
\eeq
It turns out that for any choice of the parameters $k$, $l$, and $\psi$ there exists a unique solution $(y,\hy)$ to the equations (\ref{BP_invariance_spin_transl_with_yhy}), which might not be obvious at first sight; a proof of this existence and uniqueness is provided in Appendix~\ref{app_uniqueness_RS}.

The thermodynamic aspects of the probability law (\ref{eq:measure_withbiasdistance1}) are described by its free-entropy $\ln Z$ and its Shannon entropy. In the large size limit these are extensive self-averaging quantities, we thus define the typical value of their densities as
\begin{equation}
\phi = \lim_{N \to \infty} \frac{1}{N} \mathbb{E}[ \ln Z ] \ , \qquad
s = \lim_{N \to \infty} \frac{1}{N} \mathbb{E} \left[ - \sum_{\us} \rho(\us) \ln \rho(\us) \right]  \ ,
\end{equation}
where the average $\mathbb{E}[ \bullet]$ is over the uniform choice of the $k$-uniform $l+1$-regular hypergraph $G$. The RS cavity method prediction for these quantities is obtained through the Bethe-Peierls approximation of $\ln Z$ in terms of the BP messages~\cite{KschischangFrey01,YedidiaFreeman03,MezardParisi01,MezardMontanari07}; on the translationally invariant solution this yields after a short, standard computation:
\begin{align}
\phi = & \left( 1 - \frac{(l+1)(k-1)}{k}\right) \ln 2 
- (l+1) \ln \left(1 + \frac{1}{y \hy} \right) \nonumber \\
& + \ln \left( \sum_{p=0}^{l+1}  \binom{l+1}{p} \psi(p) \hy^{-p} \right)
+ \frac{l+1}{k} \ln \left( 2^{k-1} -k-1 + \frac{k}{y} \right) \ , \\
s = & \phi - \frac{\underset{p=0}{\overset{l+1}{\sum}}  \binom{l+1}{p} \psi(p) \hy^{-p} \ln \psi(p) }{\underset{p=0}{\overset{l+1}{\sum}}  \binom{l+1}{p} \psi(p) \hy^{-p}} \ .
\label{eq_s_RS}
\end{align}
Note that for some choices of the parameters, in particular when $l$ gets large enough, this expression of the entropy $s$ becomes negative. This is impossible for a model with discrete degrees of freedom, the Shannon entropy being always non-negative, such a negativity of $s$ is thus a clear evidence of the failure of the RS hypothesis. This is however not the only mechanism for the appearance of Replica Symmetry Breaking (RSB), as we shall see next this phenomenon can occur in a phase with $s>0$.

\section{The dynamic transition}
\label{sec_dynamic}
\subsection{The reconstruction problem and its recursive distributional equations}

\label{subsec:general_setting}

We shall now present the formalism that allows to compute the location of the dynamic transition which, as explained in the introduction, manifests itself in different ways. Here we shall exploit its definition in terms of the existence of long-range point-to-set correlations in the probability measure $\rho$~\cite{MezardMontanari06,MontanariSemerjian06b}, that are related to the solvability of a tree reconstruction problem~\cite{MoPe03}. 

\begin{figure}
\begin{center}
\includegraphics[width=8.0cm]{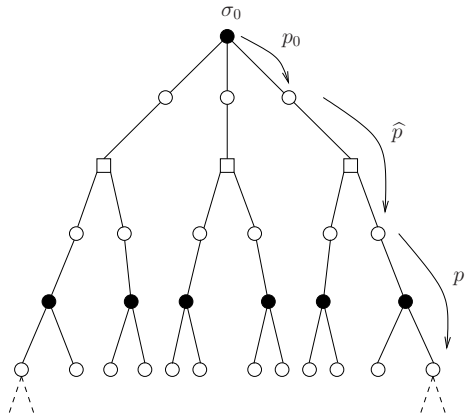}
\caption{The tree structure considered for the computation of $C_n$, represented here for $k=l+1=3$. The generation of a configuration from the law $\rho$ is performed in a broadcast fashion, the root $\s_0$ being $\pm 1$ with probability $1/2$, this information is then propagated down the tree with transmission channels $p_0$, $\hp$ and $p$.}
\label{fig:broadcast_p0_hp_p}
\end{center}
\end{figure}

Let us define the point-to-set correlation function, or overlap, at distance $n$, as follows:
\beq
C_n=\lim_{N\to \infty} \mathbb{E}[ \langle \s_0 \langle \s_0 \rangle_{\us_{B_n}} \rangle - \langle \s_0 \rangle^2 ] \ ,
\eeq
where $0$ is an arbitrary reference vertex and $B_n$ the vertices at distance at least $n$ from $0$; $\langle  \cdot \rangle$ denotes the expectation with respect to $\rho$, while $\langle  \cdot \rangle_{\us_{B_n}}$ is the conditional average with the law $\rho(\cdot | \us_{B_n})$. Note that the second term in $C_n$ actually vanishes thanks to the invariance of $\rho$ under the spin-flip transformation. The function $C_n$ can be interpreted as a measure of the correlation between the variable at the point $0$ and those in the set $B_n$, which explains its name. Because the interactions in the biased measure couple spins belonging to neighboring hyperedges it is not enough to take for $B_n$ the set of variables at distance exactly $n$ from the root; it is however equivalent to include in $B_n$ all the variables are distance at least $n$, or at distances $n$ and $n+1$. The dynamic transition separates an underconstrained (Replica Symmetric, RS) regime in which $C_n \to 0$ as $n \to \infty$, and an overconstrained (Replica Symmetry Breaking, RSB) one in which $C_n$ remains strictly positive at large distances. To compute $C_n$ we first remark that the local neighborhood of the vertex $0$, up to any finite distance, converges when $N \to \infty$ to a regular tree structure, as represented in Fig.~\ref{fig:broadcast_p0_hp_p}. Moreover the marginal law of $\rho$ on any finite neighborhood of $G$ converges, within the hypothesis of the RS solution described in Sec.~\ref{sec_RS_solution}, to a measure that admits an explicit description in terms of a broadcast process. Generating a configuration with the law of $\rho$ in a finite neighborhood of a root vertex $0$ amounts indeed to (see Fig.~\ref{fig:broadcast_p0_hp_p} for a graphical representation):
\begin{itemize}
\item choose  $\s_0=\pm 1$ with equal probability $1/2$.
\item draw the $l+1$ variables $v_1=(\s_1,w_1),\dots, v_{l+1}=(\s_{l+1},w_{l+1})$ adjacent to the root with the probability
\beq
\label{eq:channel_p0}
p_0(v_1,\dots,v_{l+1}|\s_0) = \frac{\psi\left( \underset{i=1}{\overset{l+1}{\sum}}  w_i\right)  \underset{i=1}{\overset{l+1}{\prod}}  \heta_*(w_i)}{\underset{w'_1,\dots w'_{l+1}}{\sum} \psi\left(\underset{i=1}{\overset{l+1}{\sum}}   w'_i\right) \underset{i=1}{\overset{l+1}{\prod}}   \heta_*(w'_i)}\prod_{i=1}^{l+1}\ind[\s_i=\s_0] \ .
\eeq
\item consider each of the $v_1,\dots, v_{l+1}$ variables of the first generation as the root of the subtree lying below it, and draw the value of the descendents $v_1,\dots v_{k-1}$ of a variable equal to $v$ from the conditional law
\beq
\label{eq:channel_hp}
\hp(v_1,\dots , v_{k-1}|v) = \frac{\w(\s,\s_1\dots\s_{k-1})\ind[w=\ind[\{\s_i\} \mae]] \underset{i=1}{\overset{k-1}{\prod}} \eta_*(w_i)\ind[w_i=\ind[\s,\{\s_j\}_{j \neq i} \mae]]}{\underset{\{\s'_i,w'_i\}}{\sum}\w(\s,\s'_1\dots\s'_{k-1})\ind[w=\ind[\{\s'_i\} \mae]]\underset{i=1}{\overset{k-1}{\prod}} \eta_*(w'_i)\ind[w'_i=\ind[\s,\{\s'_j\}_{j \neq i} \mae]]} \ .
\eeq
\item consider again the variables of the second generation as roots, and extract the value of their descendents from the conditional law
\beq
\label{eq:channel_p}
p(v_1 , \dots , v_l|v) = \frac{\psi\left(w+ \underset{i=1}{\overset{l}{\sum}}  w_i\right) \underset{i=1}{\overset{l}{\prod}} \heta_*(w_i)}{\underset{w'_1,\dots w'_l}{\sum}\psi\left(w+\underset{i=1}{\overset{l}{\sum}}  w'_i\right) \underset{i=1}{\overset{l}{\prod}}  \heta_*(w'_i)}\prod_{i=1}^{l}\ind[\s_i=\s] \ .
\eeq
\item iterate the last two steps until all the variables in the target neighborhood have been assigned.
\end{itemize}

This broadcast procedure, that must be performed on the $v$ variables and not only on the $\s$'s to preserve the Markov structure of the tree, can be interpreted as the transmission of an information (the value $\s_0$ at the root) through noisy channels (the conditional laws $p_0,\hp$ and $p$ defined in (\ref{eq:channel_p0},\ref{eq:channel_hp},\ref{eq:channel_p})). The question raised in the tree reconstruction problem~\cite{MoPe03} is whether the variables $\us_{B_n}$, in the large $n$ limit, contains some information on the value of $\s_0$, in the sense that the observation of $\us_{B_n}$ allows to infer the value of the root $\s_0$ with a success probability larger than the one expected from a random guess. In this Bayesian setting the optimal inference strategy is to compute the posterior probability of $\s_0$ given $\us_{B_n}$, which for Ising variables is completely described by the conditional magnetization $\langle \s_0 \rangle_{\us_{B_n}} $. The correlation function $C_n$ is a possible way of quantifying this amount of information, the tree reconstruction problem being solvable if and only if $C_n$ remains strictly positive in the large $n$ limit.

To complete the computation of $C_n$ we shall exploit again the recursive nature of the tree, but now in the opposite direction with respect to the broadcast, namely from the variables at distance $n$ towards the root. Indeed the measure $\rho(\cdot | \us_{B_n})$ is exactly described in terms of the solution of the BP equations (\ref{eq:BP_f_g}), supplemented with the boundary condition $\heta_{i\to a}(v)=\delta_{v,v_{(i,a)}}$ on the edges at distance larger than $n$ of the root, $v_{(i,a)}$ being the value taken by the variable during the broadcast. These BP messages, directed towards the root, are thus random variables because of the randomness in the boundary condition $\us_{B_n}$; one can nevertheless write recursion equations on their distributions, their law depending only on their distance from the boundary. We shall denote $P_{v,n}(\eta)$ the law of the message $\eta$ on an edge at distance $n$ from the boundary, conditional on the value of the variable on this edge being $v$ in the broadcast, and similarly $\hP_{v,n}(\heta)$ for the law of the $\heta$ messages. Putting together all the above observations leads to the following recursion equations:
\begin{align}
P_{v,n+1}(\eta) &= \sum_{v_1, \dots , v_l} p(v_1 , \dots , v_l|v) \int \prod_{i=1}^l \dd\hP_{v_i,n+1}(\heta_i) \, \delta(\eta-f(\heta_1 , \dots , \heta_l)) \ , \label{eq:1RSB_x1_P} \\
\hP_{v,n+1}(\heta) &= \sum_{v_1 , \dots , v_{k-1}} \hp(v_1 , \dots , v_{k-1}|v) \int \prod_{i=1}^{k-1} \dd P_{v_i,n}(\eta_i) \, \delta(\heta-g(\eta_1 , \dots , \eta_{k-1})) \ , \label{eq:1RSB_x1_hP}
\end{align}
where $f$, $g$, $\hp$ and $p$ have been defined in (\ref{eq:BP_eta},\ref{eq:BP_heta},\ref{eq:channel_hp},\ref{eq:channel_p}), respectively, and with the initial condition for $n=0$:
\beq
\label{eq:initialcondition_hP}
\hP_{v,0}(\heta)=\delta(\heta(\cdot)-\delta_{v,\cdot}) \ .
\eeq
The point-to-set correlation function is then computed as
\beq
 C_n= \frac{1}{2} \sum_{\s_0}\sum_{v_1 , \dots , v_{l+1}} p_0(v_1 , \dots , v_{l+1}|\s_0)\int \prod_{i=1}^{l+1}\dd \hP_{v_i,n}(\heta_i) \, \s_0 \, m(\heta_1 , \dots , \heta_{l+1}) \ ,
\label{eq:expression_CL}
\eeq
where $p_0$ is the law defined in (\ref{eq:channel_p0}), and with the expression
\begin{align}
m(\heta_1,\dots,\heta_{l+1}) &= \frac{m_+(\heta_1,\dots,\heta_{l+1}) - m_-(\heta_1,\dots,\heta_{l+1})}{m_+(\heta_1,\dots,\heta_{l+1}) + m_-(\heta_1,\dots,\heta_{l+1})} \ , \\
m_\s (\heta_1,\dots,\heta_{l+1}) &= \underset{w_1 , \dots , w_{l+1}}{\sum} \psi\left(\underset{i=1}{\overset{l+1}{\sum}} w_i \right) \underset{i=1}{\overset{l+1}{\prod}} \heta_i(\s,w_i) \ ,
\end{align}
for the conditional magnetization of the root.

The recursion equations (\ref{eq:1RSB_x1_P},\ref{eq:1RSB_x1_hP}), which are equivalent to the 1RSB equations with Parisi breaking parameter equal to $1$~\cite{MezardMontanari06}, always admit the trivial solution $P_v(\eta)=\delta(\eta-\eta_*)$, $\hP_v(\heta)=\delta(\heta-\heta_*)$ as a stationary fixed point. In the non-reconstructible (RS) phase this is the limit reached by $P_{v,n}$ and $\hP_{v,n}$ in the large $n$ limit, and then $C_n \to 0$. On the contrary in the reconstructible (RSB) phase the limit of $P_{v,n}$ and $\hP_{v,n}$ is a non-trivial fixed point, and $C_n$ remains strictly positive. For a given choice of the parameters $k$ and $\psi$ we define the dynamic transition $l_{\rm d}$ as the threshold separating these two behaviors. As $l$ is here an integer parameter we will say more precisely that $l<l_{\rm d}$ is the RS phase, $l \ge l_{\rm d}$ the RSB phase, i.e. $l_{\rm d}(k,\psi)$ is the smallest integer value of $l$ such that RSB occurs.

\subsection{Simplifications and symmetries}

The recursion equations (\ref{eq:1RSB_x1_P},\ref{eq:1RSB_x1_hP}) bear, for each value of $n$, on eight distributions $P_{v,n}$, $\hP_{v,n}$, as the variable $v=(\s,w)$ takes four different values. This number can however be divided by two thanks to the invariance of the problem under the spin-flip symmetry $\us \to -\us$. To state its consequences let us define the flip transformation $\eta \to \eta^f$ between messages, according to $\eta^f(\s,w)=\eta(-\s,w)$ (and similarly $\heta^f(\s,w)=\heta(-\s,w)$). The channels $p$ and $\hp$ being invariant under a global spin-flip, one can check that
\beq
P_{(-,w),n}(\eta)=P_{(+,w),n}(\eta^f) \ , \quad \hP_{(-,w),n}(\heta)=\hP_{(+,w),n}(\heta^f) \ , 
\eeq
which allows to close (\ref{eq:1RSB_x1_P},\ref{eq:1RSB_x1_hP}) on the four distributions $\{P_{(+,w),n},\hP_{(+,w),n}\}_{w=0,1}$, that we shall denote for simplicity $\{P_{w,n},\hP_{w,n}\}_{w=0,1}$. Using this property, as well as the invariance of $f$, $g$ under a permutation of their arguments and a more explicit version of the expressions (\ref{eq:channel_hp},\ref{eq:channel_p}) of $\hp$ and $p$, one can simplify (\ref{eq:1RSB_x1_P},\ref{eq:1RSB_x1_hP}) into:
\beq
\label{eq:recurs_P_z}
P_{w,n+1}(\eta) = \sum_{p=0}^l \frac{\binom{l}{p} \psi(p+w) \hy^{-p}}{\underset{p'=0}{\overset{l}{\sum}} \binom{l}{p'} \psi(p'+w) \hy^{-p'}} \int \prod_{i=1}^p \dd \hP_{1,n+1}(\heta_i) \prod_{i=p+1}^l \dd \hP_{0,n+1}(\heta_i)\, \delta(\eta-f(\heta_1 , \dots , \heta_l)) \ ,
\eeq
\begin{align}
\label{eq:recurs_hP_1}
\hP_{1,n+1}(\heta) &= \int \prod_{i=1}^{k-1}\dd P_{0,n}(\eta_i) \, \delta(\heta-g(\eta^f_1 , \dots , \eta^f_{k-1})) \ , \\
\label{eq:recurs_hP_0}
\hP_{0,n+1}(\heta) &= \frac{k-1}{y\hy}\int \dd P_{1,n}(\eta_1) \prod_{i=2}^{k-1}\dd P_{0,n}(\eta_i) 
\, \delta(\heta-g(\eta_1^f , \eta_2 , \dots , \eta_{k-1})) \\
\nonumber
&+ \frac{1}{\hy}\sum_{t=2}^{k-2} \binom{k-1}{t} \int \prod_{i=1}^{k-1}\dd P_{0,n}(\eta_i) \, \delta(\heta-g(\eta_1^f\dots\eta_t^f,\eta_{t+1} , \dots , \eta_{k-1})) \ .
\end{align}
The expression (\ref{eq:expression_CL}) of the correlation function $C_n$ can similarly be rewritten as:
\beq
\label{eq:real_overlap}
C_n=\sum_{p=0}^{l+1}\frac{ \binom{l+1}{p} \psi(p) \hy^{-p}}{\underset{p'=0}{\overset{l+1}{\sum}} \binom{l+1}{p'} \psi(p') \hy^{-p'}}\int\prod_{i=1}^p\dd \hP_{1,n}(\heta_i)\prod_{i=p+1}^{l+1}\dd \hP_{0,n}(\heta_i)
\, m(\heta_1,\dots,\heta_{l+1}) \ .
\eeq

A further symmetry constrains the distributions $P_{v,n}$; to unveil it let us call $P_n(\eta)$ the distribution of $\eta$ in a broadcast process wich is not conditioned on the value of the root, i.e.:
\beq
P_n(\eta)=\sum_{v=(\s,w)} \eta_*(w) P_{v,n}(\eta) \ ,
\eeq
where $\eta_*$ is normalized in such a way that $\eta_*(0)+\eta_*(1)=1/2$. Applying Bayes theorem to express the joint law of the variable at the root and those at the boundary one obtains ~\cite{MezardMontanari06}
\beq
 P_{v,n}(\eta)=\frac{\eta(v)}{\eta_*(w)} P_n(\eta) \ . 
\label{eq_Bayes}
\eeq
This yields a relation between $P_{w,n}=P_{(+,w),n}$ for the two values of $w$, namely
\beq
P_{1,n}(\eta)=y\frac{\eta(+,1)}{\eta(+,0)}P_{0,n}(\eta) \ ,
\label{eq:symmetry_P}
\eeq
where we recall that $y=\eta_*(0) / \eta_*(1) $ was defined in Eq.~(\ref{eq_def_y}). Moreover the spin-flip symmetry implies the invariance of $P_n$, i.e. $P_n(\eta) = P_n(\eta^f) $. This property, combined with (\ref{eq_Bayes}), allows to relate $P_{w,n}$ in $\eta$ and $\eta^f$ through a change of density, namely
\beq
\label{eq:symmetry_Pf}
P_{0,n}(\eta^f)=\frac{\eta(-,0)}{\eta(+,0)}P_{0,n}(\eta) \ , \qquad
P_{1,n}(\eta^f)=\frac{\eta(-,1)}{\eta(+,1)}P_{1,n}(\eta) \ .
\eeq
These symmetry relations, as well as the similar ones that hold for $\hP_{w,n}$ modulo the replacement of $y$ by $\hy$ in (\ref{eq:symmetry_P}), will be particularly useful in the treatment of the large $k$ limit presented in Sec.~\ref{sec_largek}. They imply a variety of identities between average observables, and in particular they can be used to rewrite the correlation function as
\beq
C_n=\sum_{p=0}^{l+1}\frac{ \binom{l+1}{p} \psi(p) \hy^{-p}}{\underset{p'=0}{\overset{l+1}{\sum}} \binom{l+1}{p'} \psi(p') \hy^{-p'}}\int\prod_{i=1}^p\dd \hP_{1,n}(\heta_i)\prod_{i=p+1}^{l+1}\dd \hP_{0,n}(\heta_i)
\, m(\heta_1,\dots,\heta_{l+1})^2 \ ,
\label{eq_Cn_square}
\eeq
which obviously shows that $C_n \ge 0$. This alternative form of $C_n$ can be derived by first checking that
\begin{align}
& \sum_{p=0}^{l+1} \binom{l+1}{p} \psi(p) \hy^{-p} \int\prod_{i=1}^p\dd \hP_{1,n}(\heta_i)\prod_{i=p+1}^{l+1}\dd \hP_{0,n}(\heta_i) \, A(\heta_1,\dots,\heta_{l+1})  \\  = &
\sum_{p=0}^{l+1} \binom{l+1}{p} \psi(p) \hy^{-p} \int\prod_{i=1}^p\dd \hP_{1,n}(\heta_i)\prod_{i=p+1}^{l+1}\dd \hP_{0,n}(\heta_i) \, \frac{m_-(\heta_1,\dots,\heta_{l+1})}{m_+(\heta_1,\dots,\heta_{l+1})} A(\heta_1^f,\dots,\heta_{l+1}^f)
\end{align}
for an arbitrary function $A$ which is invariant under the permutation of its arguments, and such that the integrals are well-defined, and then applying this identity with the test function $A=m(1-m)$.

\subsection{Hard fields}
\label{sec_hf}

The tree reconstruction problem considered above asks whether the observation of $\us_{B_n}$ gives some information on the value of the root $\s_0$, as quantified by the correlation function $C_n$; answering this question requires to solve the functional recursion relations (\ref{eq:recurs_P_z}-\ref{eq:recurs_hP_0}). We shall now consider a more drastic question, namely whether $\us_{B_n}$ allows to infer $\s_0$ with perfect certainty, and call $H_n$ the probability of this event. It turns out that $H_n$ is much simpler to compute than $C_n$, with scalar recursions instead of functional ones, and that $H_n$ is a lower bound for $C_n$; this last fact is quite intuitive, if $\us_{B_n}$ implies the value of $\s_0$ it certainly conveys information about it.

To explain the computation of $H_n$ let us first remark that $\s_0$ is implied by $\us_{B_n}$ if and only if all the proper bicolorings of the tree that coincides with $\us_{B_n}$ on the boundary take the same value at the root; by definition we only consider biased measures that do not strictly forbid any solution (here $\psi(p) >0$ for all $p$), hence the certain determination of $\s_0$ can only arise from the bicoloring constraints acting on the spin variables. This observation can be turned into an algorithm, called the naive reconstruction procedure: consider all the hyperedges at the boundary, and declare them ``forcing to the value $\s$'' if their $k-1$ variables at distance $n$ from the root are all equal to $-\sigma$, and ``not forcing'' otherwise. Now the variables at distance $n-1$ are assigned the value $\s$ if at least one of their incident boundary hyperedge is forcing to this value (by construction of the broadcast process there cannot be conflicting forcings to $+$ and $-$ on the same variable), and a ``white'' value $0$ if all the hyperedges are not forcing. This process can be iterated from the boundary towards the root, hyperedges being forcing if and only if $k-1$ among their variables have been assigned the same value $+1$ or $-1$. $H_n$ is thus the probability that this successive forcing mechanism percolates from the boundary to the root, with at least one of its incident hyperedge forcing it.

To  embed the analysis of this naive reconstruction algorithm into the formalism defined above we first introduce some terminology to classify the messages $\eta,\heta$; we will say that
\begin{itemize}
\item $\eta$ is forcing to $\s=\pm 1$ iff $\eta(-\s,0)=\eta(-\s,1)=0$, $\eta(\s,0)>0$ and $\eta(\s,1)>0$.
\item $\eta$ is non-forcing  iff $\eta(\s,w) > 0$ for all $\s$ and $w$.
\item $\heta$ is forcing to $\s=\pm 1$ iff $\heta(\s,1)=1$, $\heta(\s,0)=\heta(-\s,0) =\heta(-\s,1)=0$; we write then $\heta=\heta^\s$.
\item $\heta$ is non-forcing  iff $\heta(+,0)+\heta(+,1)>0$ and $\heta(-,0)+\heta(-,1)>0$.
\end{itemize}
We will also use the term hard (resp. soft) field for the forcing (resp. non-forcing) BP messages. Inserting these definitions in the BP equations (\ref{eq:BP_eta},\ref{eq:BP_heta}) one can check the combination rules argued for above: $\eta=f(\heta_1,\dots,\heta_l)$ is forcing to $\s$ iff at least one $\heta_i$ is forcing to $\s$ and none forcing to $-\s$, $\eta$ is non-forcing otherwise. Similarly $\heta=g(\eta_1,\dots,\eta_{k-1})$ is forcing to $\s$ iff all the $\eta_i$ are forcing to $-\s$, and non-forcing otherwise.

We decompose now the distributions $P_{w,n},\hP_{w,n}$ between the contributions of the hard and of the soft fields, defining
\begin{align}
P_{w,n}(\eta)&=h_{w,n} \, R_{w,n}(\eta)+(1-h_{w,n}) \, Q_{w,n}(\eta) \ , \label{eq_P_soft} \\
\hP_{w,n}(\heta)& =\hh_{w,n}\, \delta(\heta-\heta^+)+(1-\hh_{w,n})\, \hQ_{w,n}(\heta) \ , \label{eq_hP_soft}
\end{align}
where $h,\hh \in [0,1]$ are the total weights of hard fields in the corresponding distributions, the $R$ are normalized distributions on $\eta$'s forcing to $+$, and $Q$ and $\hQ$ are probability laws supported on non-forcing messages. By construction there are no messages forcing to $-$ in $P_{w,n}=P_{(+,w),n}$.

Inserting these decompositions in the recursion equations (\ref{eq:recurs_P_z}-\ref{eq:recurs_hP_0}) we see that the evolution of the hard fields weights decouple; in particular from (\ref{eq:recurs_hP_1}) we obtain $\hh_{1,n+1}=(h_{0,n})^{k-1}$ and from (\ref{eq:recurs_hP_0}) $\hh_{0,n+1}=0$, we shall thus write more simply $\hh_n$ instead of $\hh_{1,n}$. The equation (\ref{eq:recurs_P_z}) yields
\beq
h_{w,n+1}=1-\frac{\underset{p=0}{\overset{l}{\sum}} \binom{l}{p} \psi(p+w) \hy^{-p} (1-\hh_{n+1})^p}{\underset{p=0}{\overset{l}{\sum}} \binom{l}{p} \psi(p+w)\hy^{-p}}\ ,
\label{eq_hnp1}
\eeq
the recursion can thus be closed on $h_{0,n}$ and $\hh_n$, and solved starting from the initial condition $h_{0,0}=1$. Finally $H_n$ can be read off from the expression (\ref{eq:real_overlap}) of $C_n$ by isolating the contribution with at least one forcing field $\heta$ around the root, which gives
\beq
H_n = 1-\frac{\underset{p=0}{\overset{l+1}{\sum}} \binom{l+1}{p} \psi(p)\hy^{-p} (1-\hh_n)^p}{\underset{p=0}{\overset{l+1}{\sum}} \binom{l+1}{p} \psi(p)\hy^{-p}} \ .
\label{eq_Hn}
\eeq

Depending on the choice of the parameters $(l,k,\psi)$ the sequence $h_{0,n}$ (or equivalently $H_n$) either decays to 0 or to a strictly positive fixed point. The so-called rigidity threshold $l_{\rm r}(k,\psi)$ separates these two behaviors, we define it in such a way that $H_n \to 0$ when $l < l_{\rm r}$ whereas it remains strictly positive in the large $n$ limit for $l \ge l_{\rm r}$. In this latter case there is a positive probability for the observation of a far away boundary to completely determine the root (the naive reconstruction problem is solvable), hence there is certainly information about the value of the root (the usual reconstruction problem is also solvable). This observation shows that $l_{\rm r}$ is an upperbound for the dynamic threshold, $l_{\rm d}\le l_{\rm r}$.

For future use we also give here the recursion equation for the soft fields distributions, obtained by inserting the decompositions (\ref{eq_P_soft},\ref{eq_hP_soft}) into (\ref{eq:recurs_P_z}-\ref{eq:recurs_hP_0}):
\begin{align}
Q_{w,n+1}(\eta) &= \sum_{p=0}^l\frac{\binom{l}{p} \psi(p+w)\hy^{-p}(1-\hh_{n+1})^p}{\underset{p'=0}{\overset{l}{\sum}} \binom{l}{p'} \psi(p'+w) \hy^{-p'} (1-\hh_{n+1})^{p'}}\int \prod_{i=1}^p\dd \hQ_{1,n+1}(\heta_i)\prod_{i=p+1}^l \dd\hQ_{0,n+1}(\heta_i)\, \delta(\eta-f(\heta_1,\dots,\heta_l))  \ , \label{eq:soft_Q_z} \\
\hQ_{1,n+1}(\heta) &= \sum_{u=1}^{k-1}\frac{\binom{k-1}{u} (h_{0,n})^{k-1-u}(1-h_{0,n})^u}{1-(h_{0,n})^{k-1}}\int \prod_{i=1}^u\dd Q_{0,n}(\eta_i) \prod_{i=u+1}^{k-1}\dd R_{0,n}(\eta_i)\, \delta(\heta-g(\eta_1^f,\dots,\eta_{k-1}^f)) \label{eq:soft_hQ_1} \\
\hQ_{0,n+1}(\heta) &= \hP_{0,n+1}(\heta) \ . \label{eq:soft_hQ_0} 
\end{align}
The last equation comes from the absence of hard fields in $\hP_0$, one can thus take the expression (\ref{eq:recurs_hP_0}) and insert in its right hand side the decomposition (\ref{eq_P_soft}) for $P_1$ and $P_0$ to have an equation involving only the soft fields distributions; this is relatively cumbersome notationally in general, we shall only write the corresponding equation in a special case later on. The correlation function $C_n$ can also be decomposed from (\ref{eq:real_overlap}) as the sum of $H_n$ and a soft contribution:
\beq
C_n = H_n + \sum_{p=0}^{l+1}\frac{ \binom{l+1}{p} \psi(p) \hy^{-p} (1-\hh_n)^p }{\underset{p'=0}{\overset{l+1}{\sum}} \binom{l+1}{p'} \psi(p') \hy^{-p'}}\int\prod_{i=1}^p\dd \hQ_{1,n}(\heta_i)\prod_{i=p+1}^{l+1}\dd \hQ_{0,n}(\heta_i)
\, m(\heta_1,\dots,\heta_{l+1}) \ .
\label{eq_overlap_soft}
\eeq
Exploiting the symmetry relations (\ref{eq:symmetry_P},\ref{eq:symmetry_Pf}) one can rewrite the second term in this equation with the integrand squared, exactly as we did in the expression (\ref{eq_Cn_square}) of $C_n$, which proves the bound $C_n \ge H_n$ and confirms the intuition that the reconstruction problem is solvable if the naive reconstruction is.

\subsection{The Kesten-Stigum bound}
\label{sec_KS}

The study of the naive reconstruction procedure presented above has yielded the upperbound $l_{\rm d} \le l_{\rm r}$ for the threshold of the dynamic transition. We state here another upperbound, $l_{\rm d} \le l_{\rm KS}$, that is obtained by analyzing the stability of the trivial fixed point $P_v(\eta)=\delta(\eta-\eta_*)$, $\hP_v(\heta)=\delta(\heta-\heta_*)$ under the iterations of Eqs.~(\ref{eq:1RSB_x1_P},\ref{eq:1RSB_x1_hP}). The threshold $l_{\rm KS}$ is defined in such a way that for $l<l_{\rm KS}$ (resp. $l>l_{\rm KS}$) a small perturbation around the fixed point is attenuated (resp. amplified) by the iterations; for $l>l_{\rm KS}$ the sequence $P_{v,n}$ thus converges towards a non-trivial fixed point, which justifies the bound $l_{\rm d} \le l_{\rm KS}$. In the tree reconstruction literature this is known as the Kesten-Stigum transition~\cite{MoPe03,KestenStigum66}, while equivalent computations have been performed in statistical physics under the name of local instability of the RS solution towards RSB~\cite{AlmeidaThouless78}. We shall only state the expression of $l_{\rm KS}$ and refer the reader to the literature for more details on its derivation, see for instance~\cite{RiSeZd18} for an in-depth study of this transition. Let us define $M_{v,v'}$ as the probability that in a broadcast from the channel $p$ defined in (\ref{eq:channel_p}) one of the descendent variables takes the value $v'$ if the parent is equal to $v$; in formula, $M_{v,v'} = \underset{v_2,\dots,v_l}{\sum} p(v',v_2,\dots,v_l|v) $. We define in a similar way $\hM$ using instead the conditional law $\hp$ of  (\ref{eq:channel_hp}). Writing $M$ and $\hM$ as matrices, ordering their rows and columns as $v=(+,0),(+,1),(-,0),(-,1)$, one finds after a short computation:
\beq
M=\begin{pmatrix}
\alpha_0 & 1-\alpha_0 & 0 & 0 \\
1-\alpha_1 & \alpha_1 & 0 & 0 \\
0 & 0 & \alpha_0 & 1-\alpha_0 \\
0 & 0 & 1-\alpha_1 & \alpha_1 
\end{pmatrix} \ , \qquad
\hM =\begin{pmatrix}
1-\halpha_0-\halpha_1 & 0 & \halpha_0 & \halpha_1 \\
0 & 0 & 1 & 0 \\
\halpha_0 & \halpha_1 & 1-\halpha_0-\halpha_1 & 0 \\
1 & 0 & 0 & 0
\end{pmatrix} \ ,
\eeq
where the matrix elements have the following expressions,
\beq
\alpha_0 = \frac{\underset{p=0}{\overset{l-1}{\sum}} \binom{l-1}{p} \psi(p) \hy^{-p}}{\underset{p=0}{\overset{l-1}{\sum}} \binom{l-1}{p} \psi(p) \hy^{-p} + \underset{p=0}{\overset{l-1}{\sum}} \binom{l-1}{p} \psi(p+1) \hy^{-p-1}  } \ , \qquad
\alpha_1 = \frac{\underset{p=0}{\overset{l-1}{\sum}} \binom{l-1}{p} \psi(p+2) \hy^{-p-1}}{\underset{p=0}{\overset{l-1}{\sum}} \binom{l-1}{p} \psi(p+1) \hy^{-p} + \underset{p=0}{\overset{l-1}{\sum}} \binom{l-1}{p} \psi(p+2) \hy^{-p-1}  } \ , 
\eeq
\beq
\halpha_0=\frac{2^{k-2}-2}{\hy} \ , \qquad
\halpha_1=\frac{1}{y \hy} \ . 
\eeq
We then call $\theta$ the second largest, in modulus, eigenvalue of the matrix product $M \hM$ (the largest one being the Perron eigenvalue 1 as the rows are probability vectors), and define $l_{\rm KS}$ through the relation $l_{\rm KS} (k-1) \theta^2 =1$.

\section{Finite $k$ numerical results}
\label{sec:finitek_results}

\begin{figure}
\begin{center}
\includegraphics[width=8cm]{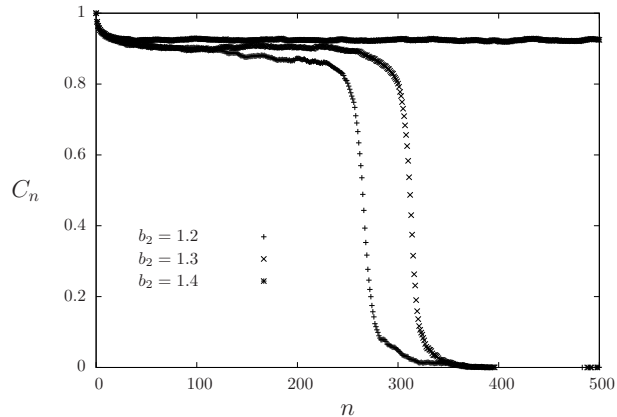}
\caption{An example of the shape of the correlation function $C_n$ as a function of $n$, here for $k=6$, $l=114$ and the bias function $\psi(p)$ defined in Eq.~(\ref{eq:bias_b1b2}), with $b_1=1.6$, $\epsilon=0.095$, and three values of $b_2$.}
\label{fig:finitek_Cvst}
\end{center}
\end{figure}

For a given choice of the parameters $(k,l,\psi)$ the model is either in a non-reconstructible, Replica Symmetric (RS) phase if the point-to-set correlation function $C_n$ decays to 0 as $n \to \infty$, or in a reconstructible, Replica Symmetry Breaking (RSB) phase if $C_n$ remains strictly positive in this large distance limit. We recall that we defined $l_{\rm d}(k,\psi)$ as the smallest integer value of $l$ which leads to RSB, and that our goal is to find a bias $\psi$ that pushes $l_{\rm d}$ to the largest possible value. We have seen several upperbounds on $l_{\rm d}$ that can be easily computed analytically: if the entropy (\ref{eq_s_RS}) computed in the RS ansatz is negative this is certainly an evidence for the RSB phenomenon; the existence of hard fields, i.e. the possibility of naive reconstruction, implies the reconstructibility, hence the rigidity bound $l_{\rm d} \le l_{\rm r}$; finally the Kesten-Stigum bound $l_{\rm d} \le l_{\rm KS}$ follows from the instability of the trivial fixed point of the reconstruction recursive equations. Nevertheless there are no lowerbounds on $l_{\rm d}$ that are simple to compute, hence an explicit determination of this threshold requires a numerical resolution of the equations (\ref{eq:recurs_P_z}-\ref{eq:recurs_hP_0}). This type of Recursive Distributional Equation (RDE) admits a natural numerical procedure to solve them, called population dynamics algorithm~\cite{ACTA73,MezardParisi01}, in which a distribution, say $P_{0,n}$, is approximated by the empirical distribution over a sample of $\N \gg 1$ representants as
\beq
P_{0,n}(\eta) \approx \frac{1}{\N} \sum_{i=1}^\N \delta(\eta-\eta_i) \ .
\eeq
With $2 \N$ fields $\eta_i$ one can encode the distributions $P_{w,n}$ at distance $n$ for $w=0,1$, from which the representants of $\hP_{w,n+1}$ are generated stochastically according to (\ref{eq:recurs_hP_1},\ref{eq:recurs_hP_0}), and in turn the populations representing $P_{w,n+1}$ can be obtained from (\ref{eq:recurs_P_z}). At each step of this iterative procedure one computes the correlation function $C_n$ from (\ref{eq:real_overlap}), interpreting the average over $\hP_{w,n}$ as an uniform sampling of an element of the corresponding population. An example of the results thus obtained is presented in Fig.~\ref{fig:finitek_Cvst}, where one sees, depending on the choices of parameters, RS cases with $C_n$ vanishing at large $n$, and RSB situations where $C_n$ remains positive. The results presented in the rest of this section have been obtained with populations of size $\N=10^6$; we considered that $C_n \to 0$ whenever the average value of $C_n$, for large enough values of $n$ such that stationarity was reached within our numerical accuracy, dropped below a small threshold value (we used $0.005$ in the figures below).

\begin{figure}
\begin{center}
\includegraphics[width=7.1cm]{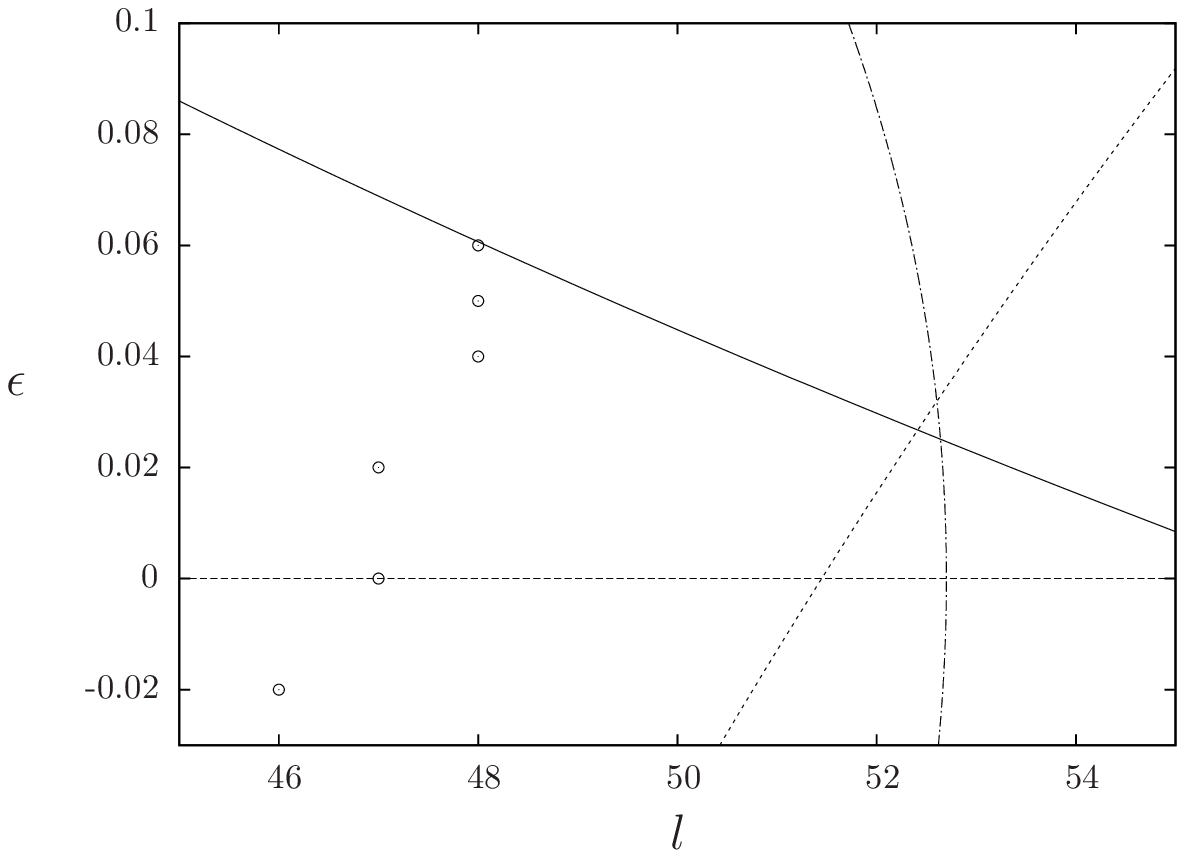}
\hspace{1.5cm}
\includegraphics[width=7.6cm]{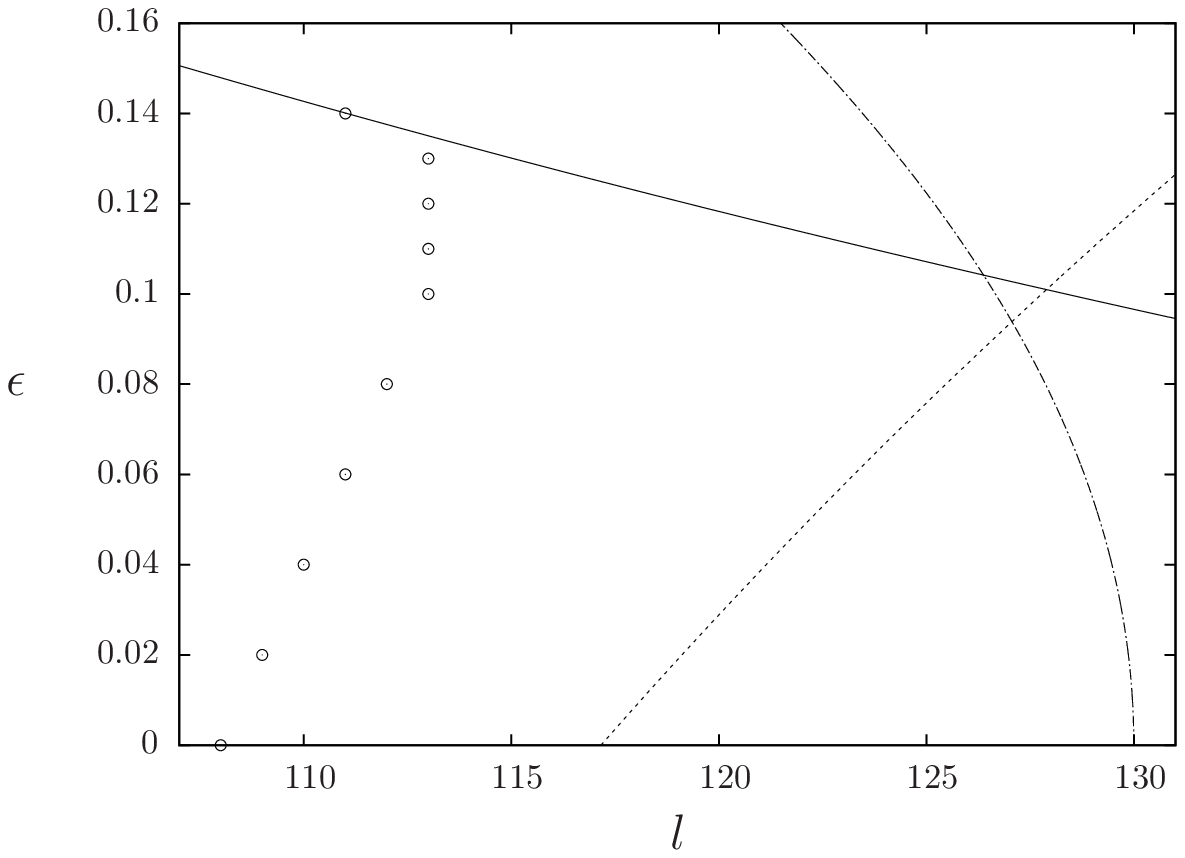}
\caption{Phase diagrams in the $(l,\epsilon)$ plane for the bias function $\psi(p)=(1-\e)^p$, for $k=5$ (left panel) and $k=6$ (right panel). The points are the clustering threshold $l_{\rm d}(\epsilon)$: for a given value of $\epsilon$ the RS phase corresponds to $l<l_{\rm d}$, the RSB phase to $l \ge l_{\rm d}$. The three continuous lines are upperbounds of $l_{\rm d}$, the area on their right is necessarily in a RSB phase; the solid one is the Kesten-Stigum bound explained in Sec.~\ref{sec_KS}, the dashed line marks the rigidity threshold $l_{\rm r}$ defined in Sec.~\ref{sec_hf}, and the dot-dashed line corresponds to the vanishing of the RS entropy (\ref{eq_s_RS}).}
\label{fig:finitek_biaisfacto}
\end{center}
\end{figure}

The function $\psi(p)$ contains a large number ($l+1$) of free parameters, some choices must hence be made on its specific form. We first considered the case where $\psi(p)=(1-\epsilon)^p$, with a single free parameter $\epsilon <1$. As explained in Sec.~\ref{sec_def_biased} this corresponds to a bias that factorizes over the hyperedges of the bicoloring problem, that we studied in~\cite{BuRiSe19} for Erd\H{o}s-R\'enyi (ER) random hypergraphs in which the degree of a vertex has a Poisson distribution of average $\alpha k$. The phase diagrams we obtained numerically for the regular case considered in this paper are presented in the $(l,\epsilon)$ parameter plane for $k=5$ and $k=6$ in Fig.~\ref{fig:finitek_biaisfacto}. They are qualitatively similar to the results obtained in~\cite{BuRiSe19} for the ER case, and quantitatively close with the correspondence $\alpha k = l+1$ between the average degree of the ER ensemble and the one fixed here. The important point we want to emphasize here is the fact that a suitable choice of $\epsilon$ allows to increase $l_{\rm d}$ with respect to its value for the uniform measure ($\epsilon=0$). For instance for $k=5$ and $l=47$, the RSB phase at $\epsilon=0$ is turned into a RS phase when $\epsilon=0.04$. Similarly for $k=6$ the dynamic transition $l_{\rm d}=108$ of the uniform measure can be pushed to $l_{\rm d}=113$ for a well-chosen value of $\epsilon$.

\begin{figure}
\begin{center}
  \includegraphics[width=7.1cm]{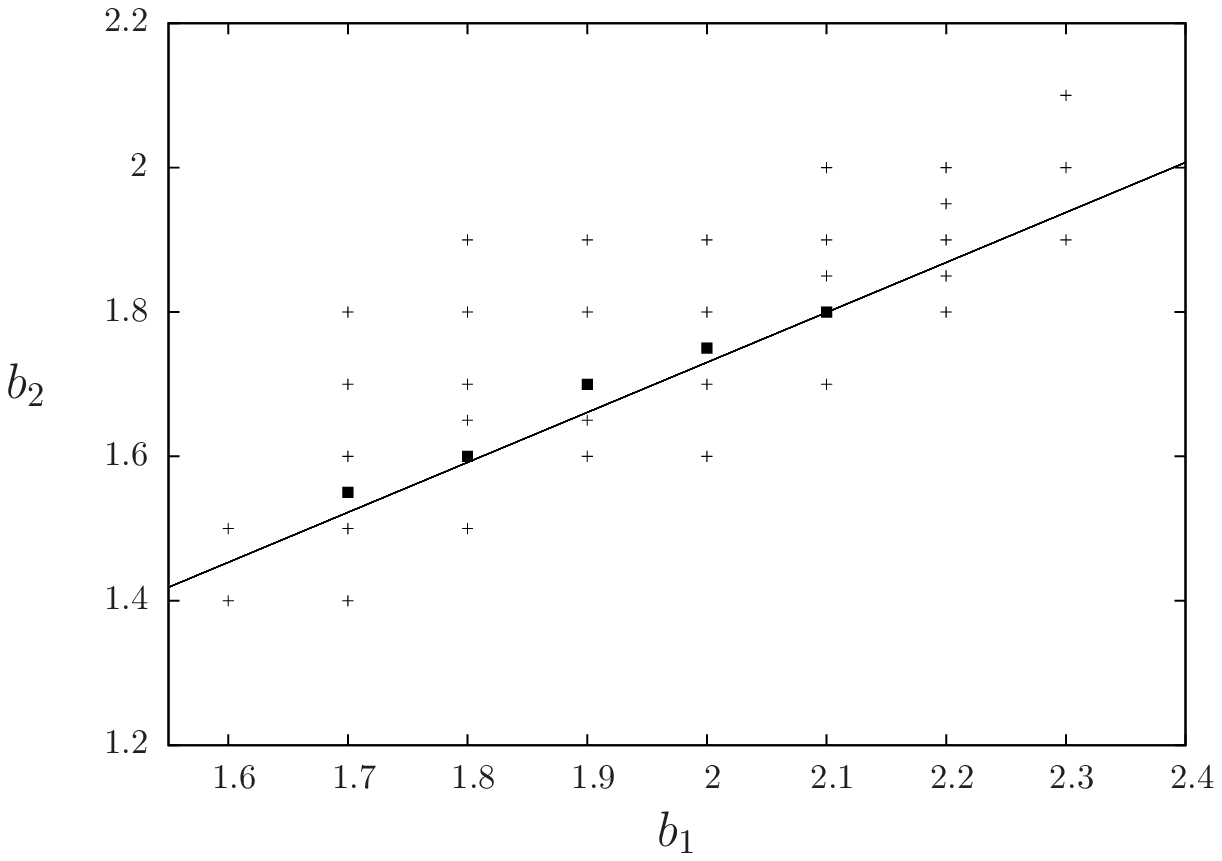}
  \hspace{5mm}
\includegraphics[width=7.1cm]{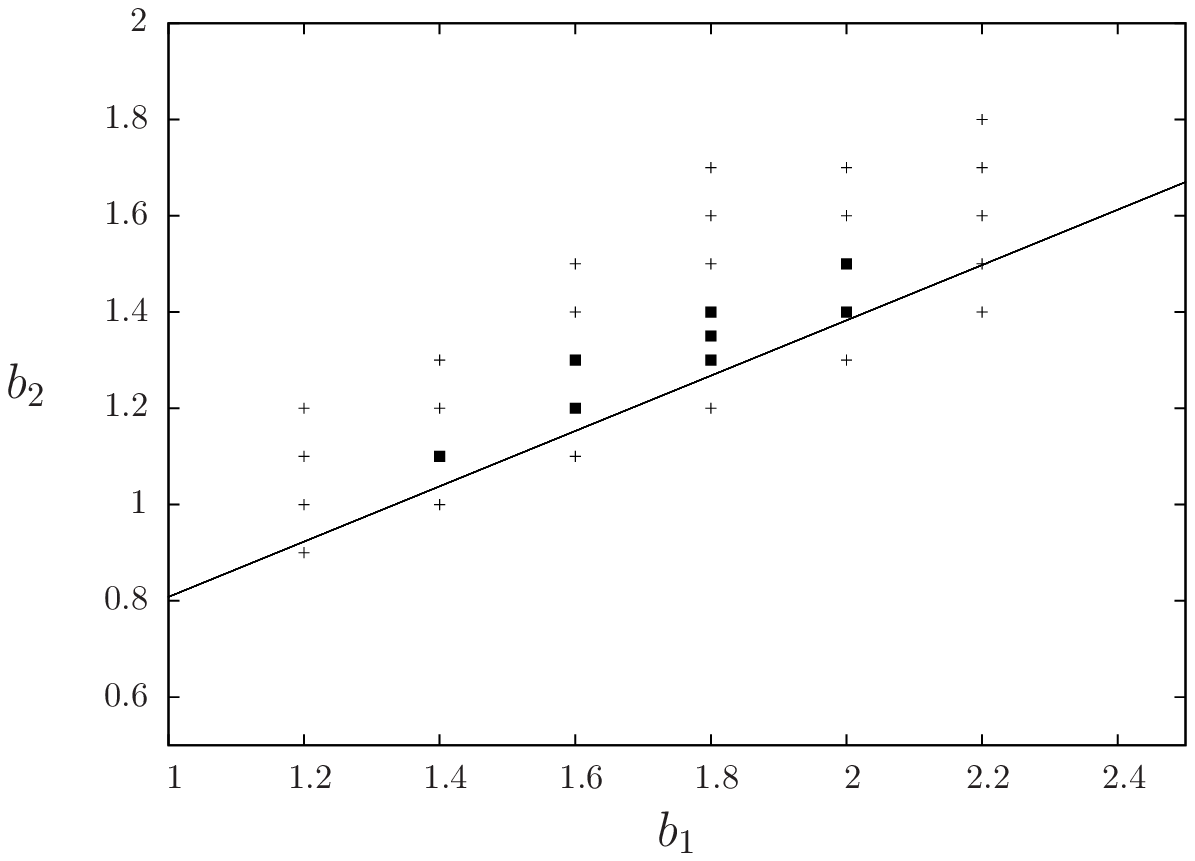}
\caption{The phase diagrams in the $(b_1,b_2)$ parameter plane for the bias function of Eq.~(\ref{eq:bias_b1b2}). The left panel is for $k=5$, $l=48$ and $\e=0.06$, the right panel for $k=6$, $l=114$, $\e=0.095$. The points marked with squares corresponds to a RS phase ($l<l_{\rm d}(k,b_1,b_2,\e)$), the crosses to a RSB phase ($l \ge l_{\rm d}(k,b_1,b_2,\e)$). The line is the Kesten-Stigum bound, the area below it is RSB.}
\label{fig:finitek}
\end{center}
\end{figure}

The natural question that arises at this point is whether the more generic bias introduced in this manuscript, i.e. the additional degrees of freedom in the choice of $\psi(p)$, allows to further increase the dynamic transition threshold $l_{\rm d}$. To investigate this point without introducing too large a space of parameters, that would be impossible to explore systematically, we considered the following function $\psi$:
\beq
\label{eq:bias_b1b2}
\psi(0)=1 \ , \quad \psi(1)=b_1 \ , \quad \psi(p)=b_2(1-\e)^p \quad \text{for} \quad p\geq 2 \ ,
\eeq
with the three free parameters $(b_1,b_2,\e)$. We then solved numerically the RDEs with parameters close to the optimal values found previously in the restricted case with $b_1=1-\epsilon$ and $b_2=1$.
The results are shown in Fig.~\ref{fig:finitek} in the parameter plane $(b_1,b_2)$, for fixed values of $k$, $l$ and $\epsilon$, with squares (resp. crosses) marking RS (resp. RSB) phases. The left panel shows the existence of a RS phase at $k=5$, $l=48$, whereas all values of $\epsilon$ led to RSB at this value of $l$ for the factorized bias $\psi(p)=(1-\e)^p$. We did not find any choice of parameters $(b_1,b_2,\e)$ with a RS phase for $l=49$. Similarly the right panel shows, for $k=6$, the largest value of $l$, $l=114$, for which we found a RS phase for well-chosen parameters (see also the drop of $C_n$ to 0 in Fig.~\ref{fig:finitek_Cvst} for $\e=0.095$, $b_1=1.6$, $b_2 = 1.2$ and $1.3$).

We summarized the main results of this Section in the Table \ref{table:result}. The first column gives the value of $l_{\rm d}$ of the uniform measure, the second column gives the result obtained with a bias of the form (\ref{eq:bias_factorized}), when optimizing on the choice of $\e$ to increase as much as possible $l_{\rm d}$, and the third column gives the result obtained with a bias of the form (\ref{eq:bias_b1b2}), for well-chosen values of $(b_1,b_2,\e)$. We can see that we were able to further improve the value of $l_{\rm d}$ for $k=5$ and $k=6$ by using a bias that introduces interactions between variables of different hyperedges, with respect to the factorized bias. An even further improvement might be achieved by a more systematic exploration of the parameter space $(b_1,b_2,\e)$, or by using even more general bias functions $\psi(p)$, at the price of a large computational cost due to the increased dimensionality of the parameter space. For comparison we give in the last column the satisfiability threshold, i.e. the smallest value of $l$ such that the typical hypergraphs have no proper bicolorings, computed within the 1RSB ansatz (see~\cite{DiSlSu13_naeksat,BrDaSeZd16} for details), that is obviously an upperbound for $l_{\rm d}$, independently of the bias. 

\begin{table}
\centering
\begin{tabular}{ | c | c | c | c | c |}
\hline
$k$ & \text{uniform} & $\e$ & $(b_1,b_2,\e)$ & $l_{\rm sat}$ \\
\hline
5 & 47 & 48 & 49 & 52 \\
\hline
6 & 108 & 113 & 115 & 129 \\
\hline
\end{tabular}
\caption{The values of $l_{\rm d}$ for the uniform measure, for the bias of equation (\ref{eq:bias_factorized}) that factorizes on the hyperedges, with the optimal value of $\epsilon$, and for the bias of equation (\ref{eq:bias_b1b2}) for  well-chosen parameters $(b_1,b_2,\e)$. The last column is the satisfiability threshold.}
\label{table:result}
\end{table}

\section{Large $k$ asymptotics}
\label{sec_largek}

The rest of the paper will be devoted to an asymptotic expansion of the clustering threshold $l_{\rm d}$ when $k\to\infty$, for the biased measure; we have just seen that for finite $k$ the latter has a larger $l_{\rm d}$ with respect to the uniform one, and that the inclusion of interactions between variables at larger distance brings a further improvement compared to a biasing function factorized over the hyperedges. It is thus natural to investigate this phenomenon in the large $k$ limit, that allows for some analytical simplifications, and where the algorithmic gap discussed in the introduction is most clearly seen. One would like in particular to understand at which order of the asymptotic expansion of $l_{\rm d}$ the effect of the bias does appear.

This Section being rather long and technical we give here, for the convenience of the reader, the main ideas and explain the organization of the forecoming computation, which is the generalization of the one we presented in~\cite{BuSe19} for the uniform measure. We will focus on the particular form of the function $\psi(p)$ defined in (\ref{eq_bias_beps}), with the two parameters $b$ and $\epsilon$, and start in Sec.~\ref{sec_largek_special} by summarizing the main equations derived above, for arbitrary $k$, in this special case. In order to take the $k\to\infty$ limit we must specify how the degree $l$ and the parameters $(b,\epsilon)$ behave with $k$; we will set
\beq
\label{eq:scaling_l}
l=2^{k-1} (\ln k + \ln\ln k + \g) \ , \qquad \epsilon = \he \, \sqrt{\frac{2}{k \ln k}} \ ,
\eeq
where $\gamma$ and $\he$ are constants independent of $k$ that parametrize the degree and the bias in this limit (the factor $\sqrt{2}$ being for later notational convenience), while $b$ will be independent of $k$. This specific choice for the scaling of $b$ and $\e$ will be justified later in this section. We will find that both $l_{\rm d}$ and the rigidity threshold $l_{\rm r}$ have asymptotic expansions of the form (\ref{eq:scaling_l}), our goal being to determine the corresponding rescaled thresholds $\g_{\rm d}$ and $\g_{\rm r}$, as a function of the parameters $(b,\he)$.

To do so we shall first expand the correlation function $C_n$ and its hard-fields contribution $H_n$, for a finite distance $n$, and find that both go to their maximal value 1, with the correction term scaling as
\beq
C_n=1-\frac{\tC_n}{k\ln k} + o\left(\frac{1}{k \ln k} \right)\ , \qquad
H_n = 1 - \frac{\tH_n}{k\ln k} +o\left(\frac{1}{k \ln k} \right) \ , 
\label{eq_scaling_Cn_Hn}
\eeq
where $\tC_n$ and $\tH_n$ are independent of $k$. These sequences depend on the rescaled parameters $\gamma$, $b$ and $\he$, and we present in Sec.~\ref{sec_largek_finite_n} recursion equations that allow to compute them (in Sec.~\ref{sec_largek_finite_n_hard} for $\tH_n$ and in Sec.~\ref{sec_largek_finite_n_soft} for $\tC_n$).

The tresholds $l_{\rm d}$ and $l_{\rm r}$ have been defined for finite $k$ according to the positivity of the large $n$ limit of the sequences $C_n$ and $H_n$, respectively. Their asymptotic expansion should thus be performed by taking the large $k$ limit after the large $n$ one; however, under the natural hypothesis (that can be checked explicitly for $H_n$) that the large $n$ limit of $C_n$ and $H_n$ is either strictly vanishing or scales with $k$ as in (\ref{eq_scaling_Cn_Hn}), one can determine $\g_{\rm d}$ and $\g_{\rm r}$ by reversing the order of the limits and studying whether $\tC_n$ and $\tH_n$ remain bounded or not in the large $n$ limit. The large $n$ limit of $\tC_n$ is thus discussed in Sec.~\ref{sec_largek_largen}. Additional difficulties need to be overcome in the intermediate regime $\g_{\rm d} < \g < \g_{\rm r}$ where reconstruction is possible but naive reconstruction is not: even if strictly hard fields are not present here the scaling (\ref{eq_scaling_Cn_Hn}) reveals that the soft fields are actually quasi-hard, the correlation function tending to one. We thus reformulate in Sec.~\ref{subsec:reweighting} the recursion of \ref{sec_largek_finite_n_soft} and put it in a form for which the large $n$ limit can be performed in a numerically tractable way. Finally our explicit results for $\g_{\rm d}$ are presented in \ref{sec_largek_results}.

\subsection{A specialization of some formulas}
\label{sec_largek_special}

Let us first specialize some of the formulas we wrote previously for a generic $\psi(p)$ to the case defined in equation (\ref{eq_bias_beps}) with the two parameters $b$ and $\epsilon$. The BP equation (\ref{eq:BP_eta}) for the function $\eta=f(\heta_1,\dots,\heta_l)$ becomes
\begin{align}
\eta(\s,1) &= \frac{1}{z} b (1-\epsilon) \prod_{i=1}^l \left(\heta_i(\s,0) + (1-\epsilon) \heta_i(\s,1)\right)\ ,\label{eq_f_beps} \\
\eta(\s,0) &= \frac{1}{z}\left[(1-b)\prod_{i=1}^l\heta_i(\s,0)  + b\prod_{i=1}^l\left(\heta_i(\s,0) + (1-\epsilon) \heta_i(\s,1)\right) \right] \ ,
\nonumber
\end{align}
for $\s=\pm 1$. The equation (\ref{BP_invariance_spin_transl_with_yhy}) for the factorized RS solution reads
\beq
y = \frac{1}{1-\epsilon} \left(1+ \frac{1-b}{b} (1+(1-\epsilon)\hy^{-1} )^{-l} \right) \ , \qquad
\hy = 2^{k-1} -k-1 + \frac{k-1}{y} \ . \label{eq_yhy_beps}
\eeq
The evolution equations (\ref{eq_hnp1},\ref{eq_Hn}) for the hard fields become
\begin{align}
h_{0,n+1}&=1-\frac{1-b+b(1+(1-\hh_{n+1})(1-\epsilon)\hy^{-1})^l}{1-b+b(1+(1-\epsilon)\hy^{-1})^l} \ , 
\label{eq:hardfields_finitek} \\
h_{1,n+1}&=1-\frac{(1+(1-\hh_{n+1})(1-\epsilon)\hy^{-1})^l}{(1+(1-\epsilon)\hy^{-1})^l} \ , 
\label{eq:hardfields_finitek2} \\
H_{n} &=1-\frac{1-b+b(1+(1-\hh_n)(1-\epsilon)\hy^{-1})^{l+1}}{1-b+b(1+(1-\epsilon)\hy^{-1})^{l+1}} \ , \label{eq:hardfields_finitek3}
\end{align}
where we recall the initial condition $h_{0,n=0}=1$ and the fact that $\hh_{n+1}=(h_{0,n})^{k-1}$. We can thus write a closed equation on $h_{0,n}$:
\begin{align}
\label{eq:h0_finitek}
h_{0,n+1}=1-\frac{1-b+b(1+(1-(h_{0,n})^{k-1}) (1-\epsilon)\hy^{-1})^l}{1-b+b(1+(1-\epsilon)\hy^{-1})^l} \ .
\end{align}
One can check numerically that this equation undergoes a discontinuous bifurcation when $l$ increases above the rigidity threshold $l_{\rm r}$. Here all the formulas depend analytically on $l$, we can thus consider it as a real parameter, even if the original model is only defined for integer $l$. The fixed point $h_0=\lim_{n\to\infty}h_{0,n}$ jumps abruptly from $0$ to a strictly positive value when $l$ is increased above $l_{\rm r}$. We can determine the location of this threshold by noting that at such a bifurcation the function that maps $h_{0,n}$ to $h_{0,n+1}$ is tangent with the diagonal, hence $l_{\rm r}$ and the bifurcating fixed point $h_{0,{\rm r}}$ are solutions of 
\begin{align}
\label{eq:rigidity_bifurcation}
h_{0,{\rm r}} &=1-\frac{1-b+b(1+(1-(h_{0,{\rm r}})^{k-1})(1-\epsilon)\hy^{-1})^{l_{\rm r}}}{1-b+b(1+(1-\epsilon)\hy^{-1})^{l_{\rm r}}} \ ,  \\  
1 & =\frac{l_{\rm r}(k-1)(h_{0,{\rm r}})^{k-2} (1-\epsilon)\hy^{-1} b(1+(1-(h_{0,{\rm r}})^{k-1}) (1-\epsilon)\hy^{-1})^{l_{\rm r}-1}}{1-b+b(1+(1-\epsilon)\hy^{-1})^{l_{\rm r}}} \ .
\end{align}
For a generic bias $\psi(p)$ the distribution $R_{w,n}$ of the hard fields introduced in (\ref{eq_P_soft}) is a priori non-trivial, but for the particular choice of $\psi$ defined in (\ref{eq_bias_beps}) it simplifies into 
\beq
R_{w,n}(\eta) = \delta(\eta-\eta^+) \ , \quad \text{where}\quad 
\eta^+(\s,0)=\frac{1}{2-\epsilon}\delta_{\s,+} \ , \quad
\eta^+(\s,1)=\frac{1-\epsilon}{2-\epsilon}\delta_{\s,+} \ ,
\label{eq_R_bias_beps}
\eeq
for all $w$ and $n$. We will also denote $\eta^-=(\eta^+)^f$ the message forcing to $-$. This allows to simplify the equation  (\ref{eq:soft_hQ_1}) on the soft fields distribution, which reads now:
\begin{align}
\hQ_{1,n+1}(\heta) &= \sum_{u=1}^{k-1}\frac{\binom{k-1}{u} (h_{0,n})^{k-1-u}(1-h_{0,n})^u}{1-(h_{0,n})^{k-1}}\int \prod_{i=1}^u\dd Q_{0,n}(\eta_i) \, \delta(\heta-g(\eta_1^f,\dots,\eta_u^f,\eta^-,\dots,\eta^-)) \ .
\label{eq:soft_hQ_1_beps} 
\end{align}
It will be useful in the following to encode in a compact way the value of $g(\eta_1,\dots,\eta_{k-1})$ when all, or almost all, the arguments of $g$ are forcing messages. We shall hence define, for a real number $\alpha$, the message $\heta=g_0(\alpha)$ as
\beq
\heta(\s,w)=\delta_{w,0}\frac{1+\sigma \tanh(\alpha)}{2} \ ;
\label{eq_def_g0}
\eeq
the value $w$ is thus fixed to $0$, while $\sigma$ can be seen as an Ising spin submitted to an effective magnetic field $\alpha$. One then founds that the values of $g$ when all its arguments are forcing are:
\begin{itemize}
\item $g(\eta^-,\dots,\eta^-)=\heta^+$ and $g(\eta^+,\dots,\eta^+)=\heta^-$, the usual combination rule to obtain a forcing message $\heta$;
\item $g(\eta^+,\eta^-,\dots,\eta^-)=g_0(\epsilon')$ and  $g(\eta^-,\eta^+,\dots,\eta^+)=g_0(-\epsilon')$, with $\epsilon'=-\frac{1}{2}\ln(1-\epsilon)$, when all the messages except one are forcing in the same direction, the last one in the opposite direction;
\item  $g(\eta^+,\dots,\eta^+,\eta^-,\dots,\eta^-)=g_0(0)$ when there are at least two forcing fields in each direction.
\end{itemize}
We will also introduce two functions $g_+$ and $g_-$ that gives the value of $g$ when all its arguments are forcing in the same direction, except one which is arbitrary, namely $g_+(\eta)=g(\eta,\eta^-,\dots,\eta^-)$ and $g_-(\eta)=g(\eta,\eta^+,\dots,\eta^+)$. Explicitly, $\heta=g_\s(\eta)$ means
\beq
\heta(\s,1)=\frac{1}{\hz}\, \eta(-\s,0) \ ,  \quad 
\heta(-\s,1)=0\ , \quad 
\heta(\s,0)=\frac{1}{\hz}\, \eta(\s,0) \ , \quad 
\heta(-\s,0)=\frac{1}{\hz}\, \eta(\s,1) \ ,
\label{eq_def_gs}
\eeq
with $\hz$ normalizing this distribution. Note that the two functions $g_+$ and $g_-$ are linked by the spin-flip operation according to $g_+(\eta^f)=(g_-(\eta))^f$.

\subsection{The large $k$ limit for finite $n$}
\label{sec_largek_finite_n}

\subsubsection{Evolution of the hard fields}
\label{sec_largek_finite_n_hard}

We start our large $k$ asymptotic expansion, using the scaling of the parameters defined in (\ref{eq:scaling_l}), by considering the solution (\ref{eq_yhy_beps}) of the translationally invariant RS equation; its leading order behavior is easily found to be
\beq
y = 1 + \he \sqrt{\frac{2}{k \ln k}} + o\left(\frac{1}{\sqrt{k\ln k}} \right)
\ , \qquad \hy = 2^{k-1} \left(1 + O\left(\frac{k}{2^{k-1}} \right) \right)  \ .
\eeq

Turning to the sequences $h_{w,n}$ for the weights of the hard fields, solutions of the recursion equations (\ref{eq:hardfields_finitek},\ref{eq:hardfields_finitek2}), one realizes easily that,  for $n$ finite in the large $k$ limit with the scaling of the parameters stated above,
\beq
\label{eq:largek_hardfields}
h_{0,n} = 1-\frac{x_{0,n}}{k\ln k} +o\left(\frac{1}{k \ln k} \right) \ ,  \quad 
h_{1,n} = 1-\frac{x_{1,n}}{k\ln k} +o\left(\frac{1}{k \ln k} \right) \ , \quad
\hh_{n+1} = 1-\frac{x_{0,n}}{\ln k} +o\left(\frac{1}{\ln k} \right) \ ,
\eeq
where $x_{0,n}$ and $x_{1,n}$ are independent of $k$ and solutions of the recursion relations:
\begin{align}
x_{0,n+1} & = B\, e^{-\g} + e^{-\g+x_{0,n}} \ , \label{eq:recurs_x0} \\
x_{1,n+1} & = e^{-\g+x_{0,n}} = x_{0,n+1}-B \, e^{-\g} \ . \label{eq:recurs_x1} 
\end{align}
Here and sometimes in the following it is more convenient to use the notation
\beq
\label{eq:def_B}
B=\frac{1-b}{b}
\eeq
as a parameter equivalent to $b$. The recursion above is closed on $x_{0,n}$, and satisfies the initial condition $x_{0,n=0} = 0$, that follows immediately from $h_{0,n=0} = 1$. Note that for $b=1$ (i.e. $B=0$) one recovers the result of equation (33) in \cite{BuSe19} for $x_n=x_{0,n}=x_{1,n}$, as it should in the uniform case. One also finds by expanding (\ref{eq:hardfields_finitek3}) that $H_n$, the hard fields contribution to the correlation function, is indeed given by the asymptotic expansion stated in (\ref{eq_scaling_Cn_Hn}), with $\tH_n=x_{0,n}$.

\begin{figure}
\begin{center}
  \includegraphics[width=7.1cm]{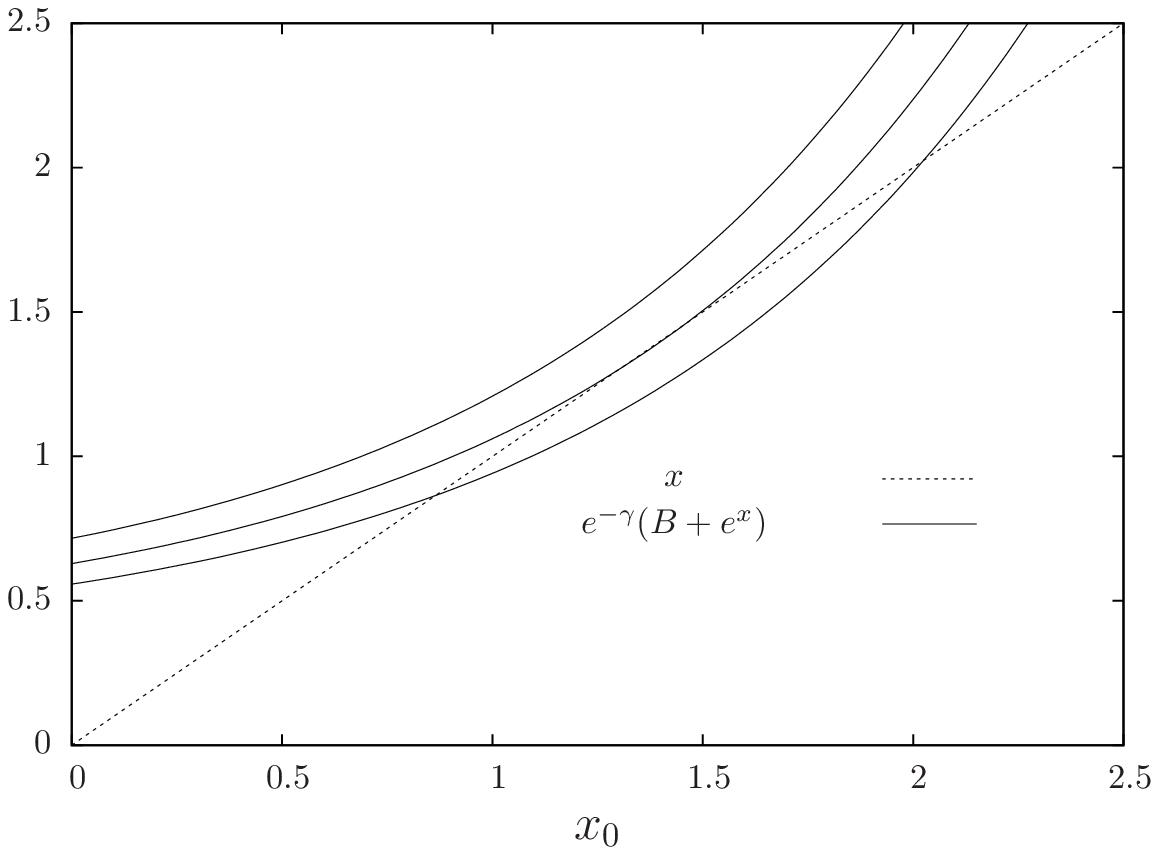}
  \hspace{5mm}
\includegraphics[width=7.1cm]{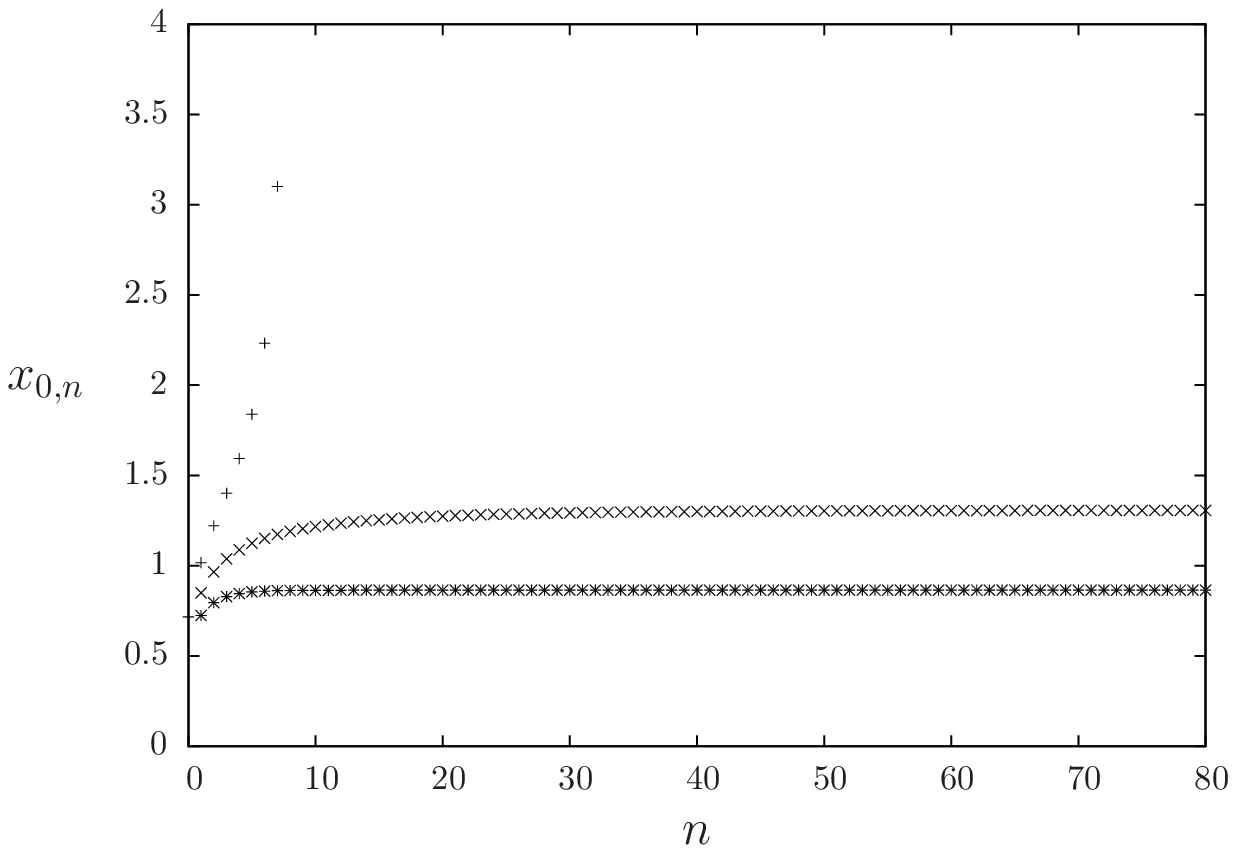}
\caption{Left panel: the functions $x$ and $B\, e^{-\g} + e^{-\g+x}$ as a function of $x$, with $b=0.4$ and from top to bottom $\g=1.25, 1.38, 1.50$. The bifurcation occurs when these two functions cross, which happens here at $\g_{\rm r}(0.4) \approx 1.378$. Right: $x_{0,n}$ as a function of $n$, for $b=0.4$ and from top to bottom $\g=1.25, 1.38, 1.50$.}
\label{fig:ath_x0vst}
\end{center}
\end{figure}

The behavior of the sequence $x_{0,n}$ solution of  (\ref{eq:recurs_x0}) is easily determined by plotting the shape of the function $x \mapsto B\, e^{-\g} + e^{-\g+x}$, see the left panel of Fig.~\ref{fig:ath_x0vst} for an example. For a given value of $b$ (hence of $B$) there exists a critical value $\g_{\rm r}(b)$ such that this function remains strictly above the diagonal when $\g < \g_{\rm r}(b)$, while it intersects it for $\g > \g_{\rm r}(b)$. As a consequence in the former case the sequence $x_{0,n}$ diverges (very rapidly, as iterated exponentials) with $n$, whereas in the latter it converges to the smallest fixed point; these behaviors are illustrated in the right panel of Fig.~\ref{fig:ath_x0vst}. The divergence of $x_{0,n}=\tH_n$ corresponds, in the large $k$ limit, to the vanishing of $H_n$ at finite $k$ (recall the definition (\ref{eq_scaling_Cn_Hn})), i.e. to the impossibility of naive reconstruction. The value of $\g_{\rm r}(b)$ can be obtained by noticing that at this bifurcation the function $x \mapsto B\, e^{-\g} + e^{-\g+x}$ is tangent with the diagonal at their unique intersection point $x_{\rm r}(b)$, hence that $(x_{\rm r}(b),\g_{\rm r}(b))$ are solution of
\beq
\label{eq:bifurc_gamma}
\begin{cases}
x= B\, e^{-\g} + e^{-\g+x} \\
1 = e^{-\g+x} 
\end{cases}
\Rightarrow \quad
\begin{cases}
x=\g \\
\g=1+B \, e^{-\g} 
\end{cases} \ .
\eeq
As $b>0$, $B>-1$, this equation admits a unique solution with $\g >0$ (the sequence $x_{0,n}$ being positive this is also the case for the fixed point $x$, and hence also of $\g$ at the bifurcation), which can be expressed as
\beq
\label{eq:gr}
\g_{\rm r}(b)=1+W\left( \frac{B}{e}\right) \ ,
\eeq
where $W(z)$ is the Lambert function, i.e. the principal solution of the equation $z=W e^W$. Note that this result coincides with the asymptotic expansion of $l_{\rm r}$ one obtains from (\ref{eq:rigidity_bifurcation}), which shows the commutativity of the limits $n \to \infty$ and $k \to \infty$ for the determination of the rigidity transition. The function $\g_{\rm r}(b)$ is plotted in the figure \ref{fig_pd_eps} (right panel, upper curve): it is a decreasing function of $b$, with $\g_{\rm r}(1)=1$ for the uniform measure. An example for the values of the fixed point reached by $x_{0,n}$ for $\g > \g_{\rm r}(b)$ can be found in the right panel of Fig.~\ref{fig:ath_tC}.

\subsubsection{Evolution of the soft fields distribution}
\label{sec_largek_finite_n_soft}

We shall now study the large $k$ limit of the soft fields distributions $Q_{w,n},\hQ_{w,n}$. The crucial point we shall exploit to simplify them is the fact that the hard fields weights $h_{w,n}$ are very close to 1 according to the scaling (\ref{eq:largek_hardfields}), hence the dominant contributions to $\hQ_{w,n}$ will arise when the incoming messages are almost all forcing. To put this remark on a quantitative ground we start with the equation (\ref{eq:soft_hQ_1_beps}) on the distribution $\hQ_{1,n+1}$. The integer $u$ that appears in this equation is a random number drawn from the binomial distribution Bin$(k-1,1-h_{0,n})$, conditioned to be strictly positive. In the large $k$ limit, using the scaling behavior (\ref{eq:largek_hardfields}) of $h_{0,n}$, one sees that the average $(k-1)(1-h_{0,n})$ of the binomial distribution vanishes as $O(1/\ln k)$, hence the main contribution in (\ref{eq:soft_hQ_1_beps}) arises from the smallest value $u=1$ appearing in the sum. We thus obtain at the leading order: 
\beq
\hQ_{1,n+1}(\heta) = \int \dd Q_{0,n}(\eta) \, \delta(\heta-g_+(\eta^f)) \ ,
\label{eq_hQ1_asymptotic}
\eeq
where the function $g_+$ was defined in (\ref{eq_def_gs}). Consider now the equation (\ref{eq:recurs_hP_0}) for $\hQ_{0,n+1}=\hP_{0,n+1}$. When all the $\eta_i$'s are extracted from the hard part of $P_{0,n}$ and $P_{1,n}$ the arguments of $g$ are all forcing, with the two possible directions represented; in most of these terms there are at least two messages in each direction, except for the term in the first line of (\ref{eq:recurs_hP_0}), and the term with $t=k-2$ in the second line. According to the discussion in Sec.~\ref{sec_largek_special} this will yield contributions of the form $g_0(0)$ and $g_0(\pm \epsilon')$, where the function $g_0$ was defined in (\ref{eq_def_g0}). When exactly one of the $\eta_i$ is a soft field, we obtain a contribution $g_-(\eta_1^f)$ from the first line if it is $\eta_1$ which is soft, a contribution $g_+(\eta_{k-1})$ from the term $t=k-2$ of the second line if $\eta_{k-1}$ is the unique soft field, all other cases leading to subdominant contributions of the form $g_0(\alpha)$ with $\alpha$ of order $\epsilon'$. Collecting these various contributions we thus obtain at the leading order: 
\begin{align}
\hQ_{0,n+1}(\heta) &= \frac{x_{1,n}}{2^{k-1}\ln k} \int \dd Q_{1,n}(\eta) \, \delta(\heta - g_-(\eta^f)) 
+ \frac{x_{0,n}}{2^{k-1}\ln k} \int \dd Q_{0,n}(\eta) \, \delta(\heta - g_+(\eta)) 
\label{eq_hQ0_asymptotic} \\
& + \beta_+ \delta(\heta-g_0(\epsilon')) + \beta_-\delta(\heta-g_0(-\epsilon')) +
\left(1-\frac{x_{0,n}+x_{1,n}}{2^{k-1}\ln k} -  \beta_+ - \beta_- \right)\delta(\heta-g_0(0)) \ ,
\nonumber
\end{align}
with
\beq
\beta_+= \frac{k-1}{\hy} (h_{0,n})^{k-1} \ , \qquad 
\beta_-=\frac{k-1}{y \hy} h_{1,n} (h_{0,n})^{k-2} \ .
\eeq

We turn now our attention to the equation (\ref{eq:soft_Q_z}) for $Q_{w,n+1}$; in the limit we are considering the non-vanishing contributions are found to arise only for values of $p$ that remain finite, the law of $p$ becoming $\cP_{w,n}(p)$, with
\beq
\cP_{1,n}(p)= \Po(p;x_{0,n})  \ , \quad
\cP_{0,n}(p)= \frac{1}{1-b+b \, e^{x_{0,n}}} \times \begin{cases}
1 & \text{if} \quad p=0 \\
b\frac{(x_{0,n})^p}{p!} & \text{if} \quad p>0
\end{cases} \ ,
\eeq
where we have introduced the notation $\Po(p;\lambda) = e^{-\lambda} \frac{\lambda^p}{p!}$ for the Poisson law of parameter $\lambda$; the $\cP_{w,n}$ are indeed well-normalized probability distributions. In the right hand side of (\ref{eq:soft_Q_z}) a (finite) number $p$ of messages $\heta_i$ are thus drawn from $\hQ_{1,n+1}$, for which we can use the limit form (\ref{eq_hQ1_asymptotic}), while the others $l-p \sim 2^{k-1} \ln k$ are drawn from $\hQ_{0,n+1}$. Observing the form of Eq.~(\ref{eq_hQ0_asymptotic}) one realizes that the number of times the first two terms of $\hQ_{0,n+1}$ will be picked become Poissonian random variables of parameter $x_{1,n}$ and $x_{0,n}$, respectively. All the other terms are of the form $g_0(\alpha)$, which will be dealt with thanks to the simple exact identity:
\beq
f(\heta_1,\dots,\heta_s,g_0(\alpha_{s+1}),\dots,g_0(\alpha_l))=f(\heta_1,\dots,\heta_s,g_0(\alpha_{s+1}+\dots+\alpha_l)) \ .
\eeq
The sum $\alpha$ of the arguments of $g_0$ is thus $\alpha=\epsilon'(a_+-a_-)$, where $a_+,a_-$ are a pair of integers drawn from the multinomial distribution of parameters $(l;\beta_+,\beta_-)$. One can thus compute the first two cumulants of $\alpha$ as 
\beq
\mathbb{E}[\alpha] = \epsilon' l (\beta_+ -\beta_-) \ , \qquad
\text{Var}[\alpha] = (\epsilon')^2 l (\beta_+ +\beta_- - (\beta_+-\beta_-)^2) \ .
\eeq
In the limit we are considering one finds that these two quantities converge to $\he^2$, while the cumulants of higher order vanish, which show that $\alpha$ tends to a Gaussian distributed random variable with mean and variance both equal to $\he^2$; we will denote the corresponding probability density as $D_{\he}\alpha = \frac{\dd \alpha}{ \sqrt{2 \pi \he^2}}e^{-\frac{1}{2 \he^2 }(\alpha - \he^2)^2}$. Note that this result justifies the choice for the scaling of $\e$ made in (\ref{eq:scaling_l}), because it leads to a finite contribution of the random variable $\alpha$, while an other scaling would have led to a trivial contribution (with mean and variance either going to $0$ or diverging with $k$). Collecting all these facts yields
\begin{align}
\label{eq:largek_ath}
Q_{w,n+1}(\eta) &= \sum_{p,q,r=0}^{\infty} \cP_{w,n}(p) \Po(q;x_{1,n}) \Po(r;x_{0,n}) \int D_{\he} \alpha \prod_{i=1}^p \dd Q_{0,n}(\eta_i) \prod_{i=p+1}^{p+q} \dd Q_{1,n}(\eta_i) \prod_{i=p+q+1}^{p+q+r} \dd Q_{0,n}(\eta_i) \\
\nonumber
&  \delta(\eta - f(g_0(\alpha),g_+(\eta_1^f),\dots ,g_+(\eta_p^f), g_+(\eta_{p+1})^f,\dots, g_+(\eta_{p+q})^f,g_+(\eta_{p+q+1}),\dots, g_+(\eta_{p+q+r}))) \ ,
\end{align}
where we used the identity $g_-(\eta^f)=g_+(\eta)^f$ to transform $q$ of the arguments of $f$.
In this equation the function $f$ is the one defined in Eq.~(\ref{eq_f_beps}), in which one can take $\epsilon=0$ at this leading order; explicitly, $\eta=f(\heta_1,\heta_2,\dots)$ means
\begin{align}
\eta(\s,1) &= \frac{1}{z} b \prod_i \left(\heta_i(\s,0) + \heta_i(\s,1)\right)\ , \label{eq_f_b_eps0} \\
\eta(\s,0) &= \frac{1}{z}\left[(1-b)\prod_i \heta_i(\s,0)  + b\prod_i \left(\heta_i(\s,0) + \heta_i(\s,1)\right) \right] \ ,
\nonumber
\end{align}
with $\s=\pm 1$.

The initial condition on $Q_{w,n=1}$ can be deduced after a short computation from the one on $\hP_{v,0}$ given in (\ref{eq:initialcondition_hP}):
\beq
\label{eq:initialcondition_Q_z}
Q_{w,1}(\eta)=\sum_{q,r=0}^{\infty} \Po(q;e^{-\g} ) \Po\left(r;\frac{e^{-\g}}{b} \right) \int D_{\he}\alpha \, \delta\left(\eta-f\left(g_0\left(\alpha+\frac{q-r}{2} \ln b\right)\right)\right) \ ,
\eeq
for both values $w=0,1$. The explicit value of $\eta$ for a given choice of $\alpha$, $q$ and $r$ reads
\beq
\eta(+,0)= \frac{1}{z} b^q e^{\alpha} \ , \quad  
\eta(+,1)= \frac{1}{z} b^{q+1} e^{\alpha} \ , \quad  
\eta(-,0)= \frac{1}{z} b^r e^{-\alpha} \ , \quad  
\eta(-,1)= \frac{1}{z} b^{r+1} e^{-\alpha} \ ,  
\eeq
with $z$ normalizing this distribution.

The recursion relation (\ref{eq:largek_ath}) bears on the two sequences of distributions $Q_{0,n}$ and $Q_{1,n}$; however the two sequences are not independent, and obey some symmetry properties, that follow from the equations (\ref{eq:symmetry_P},\ref{eq:symmetry_Pf}). In the large $k$ limit these relations translate into
\beq
\label{eq:symmetry_Q_flip}
Q_{w,n}(\eta^f)=\frac{\eta(-,w)}{\eta(+,w)}Q_{w,n}(\eta) \ , \quad \text{or equivalently} \quad \int\dd Q_{w,n}(\eta)A(\eta)=\int\dd Q_{w,n}(\eta)A(\eta^f)\frac{\eta(-,w)}{\eta(+,w)} \ ,
\eeq
and
\beq
\label{eq:symmetry_Q_01}
x_{0,n}Q_{0,n}(\eta)=\frac{\eta(+,0)}{\eta(+,1)}x_{1,n}Q_{1,n}(\eta) \ , \quad x_{0,n}\int\dd Q_{0,n}(\eta)A(\eta)=x_{1,n}\int\dd Q_{1,n}(\eta)A(\eta)\frac{\eta(+,0)}{\eta(+,1)} \ ,
\eeq
for any function $A$ such that the integrals exist. One can check by induction on $n$ that the sequences $Q_{0,n}$ and $Q_{1,n}$ solution of (\ref{eq:largek_ath}) with the initial condition (\ref{eq:initialcondition_Q_z}) do indeed satisfy these identities.

We can finally establish the scaling stated in (\ref{eq_scaling_Cn_Hn}) for the correlation function $C_n$, by simplifying the expression (\ref{eq_overlap_soft}) in the large $k$ limit. Observing in particular that the probability law for $p$ is the same in this equation and in (\ref{eq:soft_Q_z}) with $w=0$ (modulo the shift $l \to l+1$ which is irrelevant in the limit), one finds after after a short computation the expression
\beq
\label{eq:reduced_overlap_n}
\tC_n= x_{0,n}\int \dd Q_{0,n}(\eta)(1-\widetilde{m}(\eta))\ , 
\eeq
for the reduced correlation function $\tC_n$, where we defined
\beq
\widetilde{m}(\eta)=\frac{\eta(+,0)-\eta(-,0)}{\eta(+,0)+\eta(-,0)} \ .
\eeq
Note that $\tC_n$ satisfies the inequalities $0\leq\tC_n\leq x_{0,n}=\tH_n$, which are immediate consequences of the bounds $H_n \le C_n \le 1$ we obtained at finite $k$ and of the definitions in (\ref{eq_scaling_Cn_Hn}). As a consistency check one can also derive the bounds on $\tC_n$ directly in the large $k$ formalism; one of them is obvious from the observation that $\widetilde{m}(\eta) \le 1$ for all $\eta$, the other one follows from the identity
\beq
\int \dd Q_{0,n}(\eta) \, \widetilde{m}(\eta) = \int \dd Q_{0,n}(\eta) \, \widetilde{m}(\eta)^2 \ge 0 \ ,
\eeq
which can be proven from the Bayes symmetry expressed in (\ref{eq:symmetry_Q_flip}), using the test function $A(\eta) = \widetilde{m}(\eta)(1- \widetilde{m}(\eta)) $. 

\subsection{The large $n$ limit}
\label{sec_largek_largen}

Let us summarize what we have just achieved and underline the main equations that will be used in the following. We have obtained recursive equations, in which the parameter $k$ has disappeared, that allow to compute the reduced correlation function $\tC_n$ and its hard-fields contribution $\tH_n$ introduced in (\ref{eq_scaling_Cn_Hn}). The latter can be obtained from the scalar recursion (\ref{eq:recurs_x0}), it depends on $\g$ and $b$, and the asymptotic expansion of the rigidity threshold is of the form (\ref{eq:scaling_l}) with a constant $\g_{\rm r}(b)$ easily determined from the large $n$ behavior of $\tH_n$: for $\g < \g_{\rm r}(b)$ one has $\tH_n \to \infty$ as $n \to \infty$, while $\tH_n$ remains bounded for $\g \ge \g_{\rm r}(b)$. The computation of the reduced correlation function $\tC_n$ requires instead the resolution of the functional recursion equation (\ref{eq:largek_ath}) on the distributions of the soft-fields $Q_{w,n}$, supplemented by the initial condition (\ref{eq:initialcondition_Q_z}), from which $\tC_n$ is computed using the equation (\ref{eq:reduced_overlap_n}). The sequence $\tC_n$ depends on the parameters $\g$, $b$ and $\he$, and the constant $\g_{\rm d}(b,\he)$ in the asymptotic expansion of the dynamic threshold $l_{\rm d}$ is deduced from the large $n$ asymptotics of $\tC_n$ (if $\g < \g_{\rm d}(b,\he)$ then $\tC_n \to \infty$, while it remains bounded for $\g > \g_{\rm d}(b,\he)$). We shall now discuss the computation of $\tC_n$ in the large $n$ limit, as the final step to complete the determination of $\g_{\rm d}(b,\he)$.

\subsubsection{For $\g>\g_{\rm r}(b)$}

The most natural way to solve numerically the functional recursion equation (\ref{eq:largek_ath}) on $Q_{w,n}$ is to use the population dynamics algorithm already explained in Sec.~\ref{sec:finitek_results}, that consists in approximating $Q_{w,n}$ by the empirical distribution over a sample of $\cal{N}$ representative elements $\{\eta_1,\dots,\eta_{\cal{N}}\}$. An iteration step $n \to n+1$ amounts to update the populations by drawing the integers $p$, $q$, and $r$ from their respective laws, extracting the $\eta_i$'s from the current populations, and creating an $\eta$ of the new population according to the argument of the Dirac delta in (\ref{eq:largek_ath}). When $\g>\g_{\rm r}(b)$ this procedure can be performed without difficulty for arbitrarily large distances $n$, as the sequences $x_{0,n},x_{1,n}$ remain bounded for all $n$. 

\begin{figure}
\begin{center}
  \includegraphics[width=7.6cm]{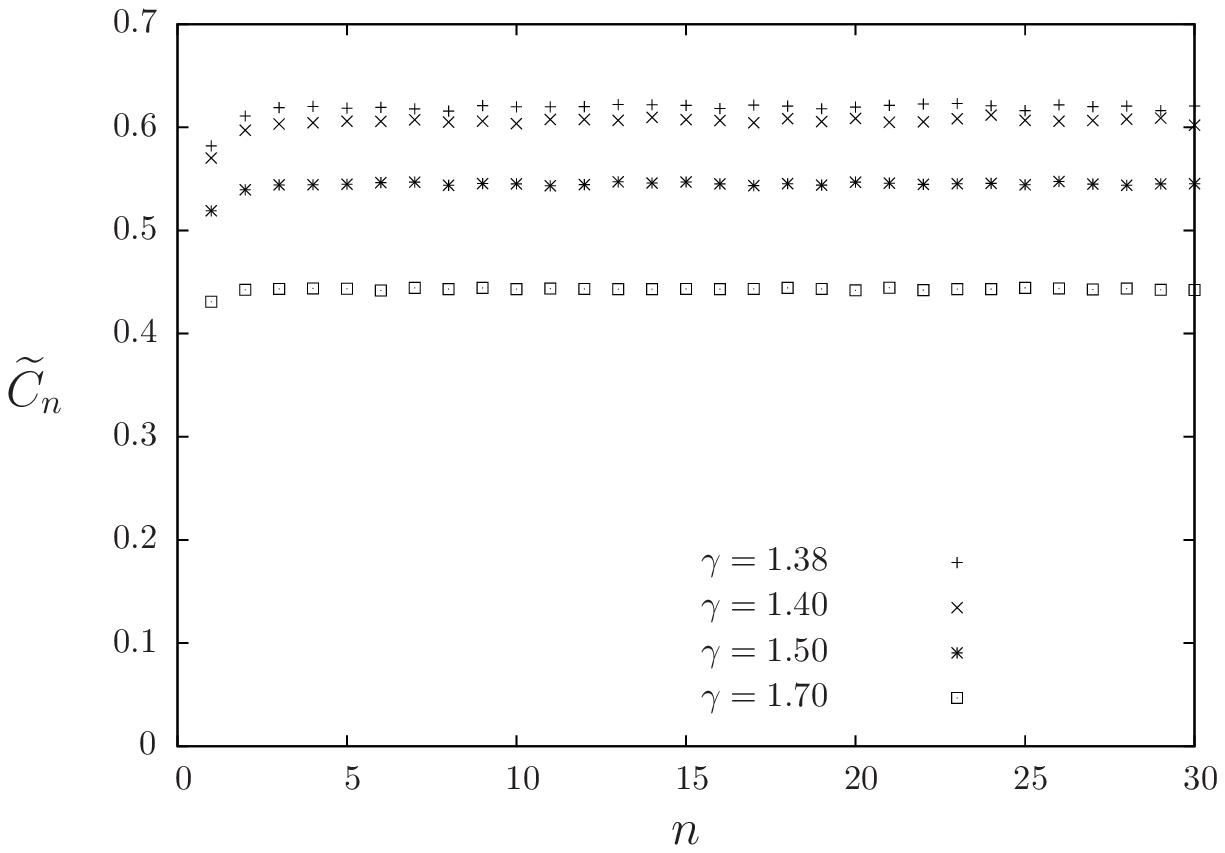}
  \hspace{1.5cm}
\includegraphics[width=7.6cm]{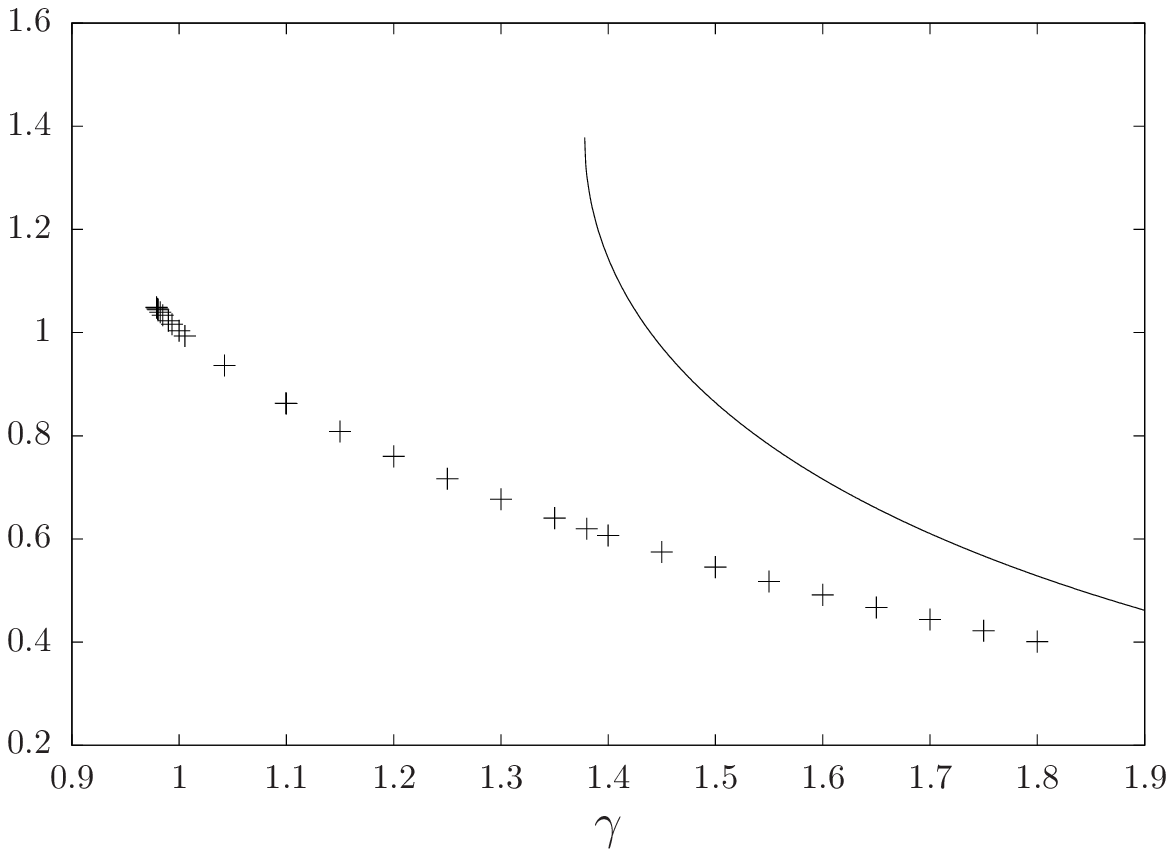}
\caption{Left: the reduced correlation function $\tC_n$ as a function of the distance $n$ for $b=0.4$, $\he=0$, and several values of $\g$ larger than the rigidity threshold $\g_{\rm r}(0.4) \approx 1.378$. Right: the large distance limit $\tC$ (points) and its hard-field contribution $\tH$ (solid line) as a function of $\g$ for $b=0.4$ and $\he=0$. The points of $\tC$ below the rigidity threshold have been obtained with the reweighted algorithm presented in Sec.~\ref{subsec:reweighting}.}
\label{fig:ath_tC}
\end{center}
\end{figure}

The figure~\ref{fig:ath_tC} presents numerical results obtained in this way for $b=0.4$ and $\he=0$. We have plotted on the left panel $\tC_n$ as a function of $n$ for some values of $\g$ above the rigidity threshold $\g_{\rm r}(b) \approx 1.378$. One can see that $\tC_n$ converges at large $n$ to a finite limit $\tC$, that we have plotted as a function of $\g$ in the right panel, along with the limit $\tH$ of $\tH_n=x_{0,n}$. As we mentioned before the reduced overlap satisfy the bounds $0\leq\tC_n\leq \tH_n$, hence in the large $n$ limit one has $0\leq\tC\leq \tH$, that is indeed verified in the right panel of the figure~\ref{fig:ath_tC}. This implies that $\tC_n$ remains bounded for $\g>\g_{\rm r}(b)$, hence the expected inequality $\g_{\rm d}(b,\he) \le \g_{\rm r}(b)$. The observation of the right panel of figure~\ref{fig:ath_tC} suggests the less obvious fact that this inequality is strict; indeed $\tH$ has a square root singularity when $\g \to \g_{\rm r}^+$, as a consequence of the bifurcation it undergoes, while $\tC$ seems pefectly smooth in this limit, suggesting that it remains finite down to a critical value $\g_{\rm d} < \g_{\rm r}$.

Unfortunately the most interesting regime $\g_{\rm d} < \g < \g_{\rm r}$ cannot be studied with the simple numerical procedure we just described: when $\g<\g_{\rm r}(b)$ the sequences $x_{0,n}$ and $x_{1,n}$ diverge, hence the random numbers $p,q,r$ of fields $\eta_i$ that must be manipulated to implement (\ref{eq:largek_ath}) become very quickly too large for any practical purpose. We shall thus devise in the next subsection an alternative formulation to circumvent this difficulty, that was used in particular to obtain the points of the curve $\tC$ below the rigidity threshold in the right panel of figure~\ref{fig:ath_tC}.

\begin{figure}
\begin{center}
\includegraphics[width=8cm]{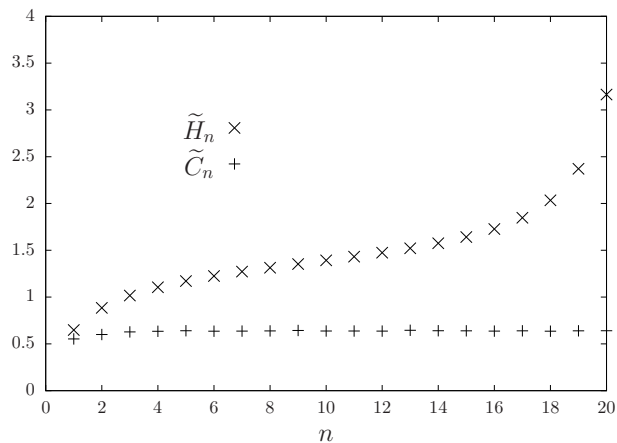}
\caption{The reduced correlation function $\tC_n$ and its hard-field contribution $\tH_n$ as a function of the distance $n$ for $b=0.4$, $\he=0$ and $\g=1.35$, slightly below the rigidity threshold.}
\label{fig:xtCvst_b0p4_g1p35}
\end{center}
\end{figure}

In order to give an intuition on how this reformulation should be performed we first present in figure~\ref{fig:xtCvst_b0p4_g1p35} the results of the simple procedure for $\g$ slightly below $\g_{\rm r}$, and distances $n$ not too large. One sees clearly in this plot that $\tH_n=x_{0,n}$ diverges, while $\tC_n$ seems to remain bounded; the expression (\ref{eq:reduced_overlap_n}) of $\tC_n$ reveals that such a situation is possible if $Q_{0,n}$ concentrates on fields $\eta$ with $\widetilde{m}(\eta)$ very close to 1. By inspection of the populations in our numerical simulations we have checked that this is indeed the case, and more precisely that both $Q_{0,n}$ and $Q_{1,n}$ tend to a Dirac peak on the hard-field $\eta^+$. The finite value of $\tC_n$ in the large $n$ limit of the intermediate regime $\g_{\rm d}(b,\he)< \g < \g_{\rm r}(b)$ which is reconstructible without strictly hard-fields arises thus from a delicate compensation in the multiplication of the diverging factor $x_{0,n}$ and of the vanishing integral $\int \dd Q_{0,n}(\eta)(1-\widetilde{m}(\eta))$. The relevant contribution of the latter arises from atypical values of $\eta$ for which $Q_{0,n}$ is of order $1/x_{0,n}$, the typical values of $\eta \approx \eta^+$ having $1-\widetilde{m}(\eta) \approx 0$.

\subsubsection{A reweighting scheme}
\label{subsec:reweighting}

To handle the difficulty that arises in the intermediate regime $\g_{\rm d} < \g < \g_{\rm r}$ we will adapt the approach we developed in~\cite{BuSe19} for the uniform measure, introducing a reweighted probability distribution $\mu_n(\eta)$ that gives less importance to the typical quasi-hard fields which do not contribute to $\tC_n$. We define it as
\beq
\label{eq:definition_mu_n}
\mu_n(\eta)=x_{1,n}Q_{1,n}(\eta) \sqrt{\frac{\eta(-,1)}{\eta(+,1)}} \ ,
\eeq 
the reweighting factor proportional to $\sqrt{\frac{\eta(-,1)}{\eta(+,1)}}$ indeed vanishes when $\eta$ is a hard-field $\eta^+$. This choice also ensures the invariance of $\mu_n$ under a spin-flip transformation, $\mu_n(\eta^f)=\mu_n(\eta)$, as can be easily seen from the first equality in (\ref{eq:symmetry_Q_flip}) with $w=1$. Note that $\mu_n$ is a positive measure, but not a normalized probability measure anymore. One can nevertheless check that its total mass, that we shall denote $m_n$, is finite for all finite $n$. One has indeed, inverting the relation (\ref{eq:definition_mu_n}) and exploiting the normalization of $Q_{1,n}$,
\beq
x_{1,n} = \int \dd \mu_n(\eta) \, \sqrt{\frac{\eta(+,1)}{\eta(-,1)}} = \int \dd \mu_n(\eta) \, \frac{1}{2} \left(\sqrt{\frac{\eta(+,1)}{\eta(-,1)}} + \sqrt{\frac{\eta(-,1)}{\eta(+,1)}} \right) = \int \dd \mu_n(\eta) \, \frac{\eta(+,1) + \eta(-,1)}{2 \sqrt{\eta(+,1) \eta(-,1)}} \ ,
\label{eq_x1n_as_mu}
\eeq
where we used the invariance $\mu_n(\eta^f)=\mu_n(\eta)$ to symmetrize the integrand. Thanks to the inequality between arithmetic and geometric means of positive numbers the last integral is larger than $m_n$, which implies finally $m_n \le x_{1,n}$. We can thus define a probability distribution $\nu_n$ by dividing $\mu_n$ by its total mass, $\nu_n(\eta) = \mu_n(\eta) / m_n$. The problem at hand is equivalently described in terms of the $\{x_{w,n} ,Q_{w,n}\}$, of $\mu_n$, or of the pair $(\nu_n,m_n)$; for instance the reduced overlap can be expressed as
\beq
\tC_n = m_n \int \dd \nu_n(\eta) C(\eta) \ , \qquad 
C(\eta)=\frac{1}{\sqrt{\eta(+,1) \eta(-,1)}} \frac{2 \eta(+,0) \eta(-,0)}{\eta(+,0) + \eta(-,0)}  \ ,
\label{eq:tC_computedfrom_nu}
\eeq
where we used (\ref{eq:symmetry_Q_01}) to transform the integral over $Q_{0,n}$ as one over $Q_{1,n}$.
As we shall see the reweighted formulation is however much more convenient to study the large $n$ limit, as it avoids the direct manipulation of the diverging quantities $x_{w,n}$.

We will now derive recursion relations for $(\nu_n,m_n)$; to do so it will be convenient to first introduce a different parametrization of the messages $\eta$. These are normalized probability distributions on a space of four states $(\s,w)$, they can be thus encoded with three real numbers, that we shall choose as
\beq
\ua= \sqrt{\frac{\eta(+,1)}{\eta(-,1)}} \ , \quad
\ub=\sqrt{\frac{\eta(+,0)-\eta(+,1)}{\eta(-,0)-\eta(-,1)}} \ , \quad
\uc=\sqrt{\frac{(\eta(+,0)-\eta(+,1)) (\eta(-,0)-\eta(-,1))}{B^2 \, \eta(+,1) \eta(-,1)}} \ .
\label{eq_def_u}
\eeq
We will group them as a row vector with three columns, $u=(\ua,\ub,\uc)$, and define for later use the associated canonical basis $\ea=(1,0,0)$, $\eb=(0,1,0)$, $\ec=(0,0,1)$. Consider now the BP equation $\eta=f(\heta_1,\heta_2,\dots)$ defined in (\ref{eq_f_b_eps0}); it becomes in terms of this parametrization
\begin{align}
\ua &= \prod_i \sqrt{\frac{\heta_i(+,0) + \heta_i(+,1)}{\heta_i(-,0) + \heta_i(-,1)}} \ , \qquad
\ub = \prod_i \sqrt{\frac{\heta_i(+,0)}{\heta_i(-,0)}} \ , \label{eq_BP_u} \\
\uc &= \prod_i \sqrt{\frac{\heta_i(+,0) \heta_i(-,0)}{(\heta_i(+,0) + \heta_i(+,1)) (\heta_i(-,0) + \heta_i(-,1))}}
\ . \nonumber
\end{align}
This shows that the arguments of the square roots in (\ref{eq_def_u}) are non-negative numbers, as they should for the definition of $u$ to be meaningful. Moreover this expression reveals the motivation for this peculiar choice of parametrization: the BP equation $\eta=f(\heta_1,\heta_2,\dots)$ becomes multiplicative with respect to its arguments when $\eta$ is expressed in terms of $u$. We will also use the notation $\tu=(\tua,\tub,\tuc)$ with $u=(e^{\tua},e^{\tub},e^{\tuc})$; as the components of $u$ are positive those of $\tu$ are real numbers, and the BP equation becomes additive in terms of $\tu$. It will also be useful to define the spin-flip operation on the triplets $u$ and $\tu$; as $\eta^f(\s,w) = \eta(-\s,w)$ one deduces easily from (\ref{eq_def_u}) the corresponding transformations:
\beq
u^f=\left( \frac{1}{\ua}, \frac{1}{\ub}, \uc \right) \ , \qquad
\tu^f=(-\tua , -\tub , \tuc ) \ .
\eeq
In the following we will take the liberty to use the three equivalent parametrizations $\eta$, $u$ and $\tu$ according to which one is the most convenient, keeping implicit the relationships between them that we have just defined.

Let us now rewrite (\ref{eq:largek_ath}) by translating the image $\eta$ of the function $f$ in the $\tu$ parametrization: 
\begin{align}
Q_{w,n+1}(\eta) &= \sum_{p,q,r=0}^{\infty} \cP_{w,n}(p) \Po(q;x_{1,n}) \Po(r;x_{0,n}) \int D_{\he} \alpha \prod_{i=1}^p \dd Q_{0,n}(\eta_i) \prod_{i=p+1}^{p+q} \dd Q_{1,n}(\eta_i) \prod_{i=p+q+1}^{p+q+r} \dd Q_{0,n}(\eta_i) 
\label{eq_Qw_tu}
\\
\nonumber
&  \delta\left(\tu - V_0(\alpha) - \sum_{i=1}^p V_+(\eta_i^f) - \sum_{i=p+1}^{p+q} V_-(\eta_i) - \sum_{i=p+q+1}^{p+q+r} V_+(\eta_i) \right) \ ,
\end{align}
where we defined
\begin{align}
V_0(\alpha) & = (\alpha,\alpha,0) \ , \\
V_+(\eta) & = \left( \frac{1}{2} \ln\left( \frac{\eta(+,0)+\eta(-,0)}{\eta(+,1)} \right) \, , 
\frac{1}{2} \ln\left(\frac{\eta(+,0)}{\eta(+,1)} \right) \, , 
\frac{1}{2} \ln\left( \frac{\eta(+,0)}{\eta(+,0)+\eta(-,0)} \right)\right) \ , \\
V_-(\eta)&=V_+(\eta)^f =\left(- \frac{1}{2} \ln\left( \frac{\eta(+,0)+\eta(-,0)}{\eta(+,1)} \right) \, , 
- \frac{1}{2} \ln\left(\frac{\eta(+,0)}{\eta(+,1)} \right) \, , 
\frac{1}{2} \ln\left( \frac{\eta(+,0)}{\eta(+,0)+\eta(-,0)} \right)\right) \ .
\end{align}
For completeness we also state the expression of $V_+$ with its argument translated in the $u$ parametrization, namely
\begin{align}
V_+^{(1)}(\eta) & = \frac{1}{2} \ln\left( 1 + B \frac{\ub \uc}{\ua} + \frac{1}{(\ua)^2} + B \frac{\uc}{\ua \ub} \right) \ , \\
V_+^{(2)}(\eta) & =  \frac{1}{2} \ln\left( 1 + B \frac{\ub \uc}{ \ua} \right) \ , \\
V_+^{(3)}(\eta) & =  V_+^{(2)}(\eta) - V_+^{(1)}(\eta) \ , \\
\end{align}
Because of the additivity property of the parametrizations in terms of $\tu$ it is easier to describe $Q_{w,n}$ in terms of its characteristic function, that we define as
\beq
\Xi_{w,n}(z)=\int \dd Q_{w,n}(\eta) \, e^{iz \cdot \tu} \ ,
\eeq
where $z=(\za,\zb,\zc)$ and we denoted the standard scalar product $z \cdot \tu = \za \tua + \zb \tub + \zc \tuc $. Indeed the equation (\ref{eq_Qw_tu}) translates into
\begin{align}
\Xi_{w,n+1}(z)= e^{\he^2 (i (\za+\zb) - \frac{1}{2} (\za+\zb)^2 )} &
\sum_{p,q,r=0}^{\infty} \cP_{w,n}(p) \Po(q;x_{1,n}) \Po(r;x_{0,n}) 
\left( \int \dd Q_{0,n}(\eta) e^{i z \cdot V_+(\eta^f) }\right)^p \\
& \left( \int \dd Q_{1,n}(\eta) e^{i z \cdot V_-(\eta) }\right)^q
\left( \int \dd Q_{0,n}(\eta) e^{i z \cdot V_+(\eta) }\right)^r \ ,
\end{align}
where the first factor comes from the Gaussian integration on $\alpha$. For $w=1$ the three integers $p,q,r$ have Poisson distributions, the sums can then be easily performed to obtain
\begin{align}
\Xi_{1,n+1}(z)&=\exp\left[ \he^2 (i (\za+\zb) - \frac{1}{2} (\za+\zb)^2 ) -x_{1,n}-2x_{0,n} +
x_{1,n}\int\dd Q_{1,n}(\eta) e^{iz \cdot V_-(\eta)} 
\right. \nonumber
\\  & \left. \qquad \qquad
+x_{0,n}\int\dd Q_{0,n}(\eta)\left(e^{i z \cdot V_+(\eta)} + e^{i z \cdot V_+(\eta^f)} \right)  \right] \ .
\label{eq_Xi1}
\end{align}

We can now come back to the reweighted measure $\mu_n$ we introduced in (\ref{eq:definition_mu_n}), and its normalized version $\nu_n$, for which we define the characteristic functions similarly
\beq
\hmu_n(z)=\int\dd\mu_n(\eta) \, e^{i z \cdot \tu} \ , \quad \hnu_n(z)=\int\dd\nu_n(\eta) \, e^{iz \cdot \tu} = \frac{1}{m_n} \hmu_n(z) \ .
\eeq
The reweighting factor $\sqrt{\frac{\eta(-,1)}{\eta(+,1)}}$ between $\mu_n$ and $Q_{1,n}$ can be expressed as $e^{-\tua}$, the characteristic functions of these two measures are thus linked by a simple shift of their arguments:
\beq
\mu_n(\eta) =x_{1,n}Q_{1,n}(\eta) e^{-\tua}  \iff \hmu_n(z)=x_{1,n}\Xi_{1,n}(z+i \ea) \ .
\eeq
Using this shift of argument in (\ref{eq_Xi1}), and recalling from (\ref{eq:recurs_x1}) that $x_{1,n+1} = e^{-\g+x_{0,n}}$
we obtain:
\begin{align}
\hmu_{n+1}(z)&=\exp\left[-\g - \frac{\he^2}{2} - \frac{\he^2}{2} (\za+\zb)^2  - x_{0,n} - x_{1,n}
+ x_{1,n}\int\dd Q_{1,n}(\eta) \sqrt{\frac{\eta(+,0)+\eta(-,0)}{\eta(+,1)}}  e^{iz \cdot V_-(\eta)} 
\right. \nonumber
\\  & \left. \qquad \qquad
 +x_{0,n}\int\dd Q_{0,n}(\eta) \left( 
\sqrt{\frac{\eta(+,1)}{\eta(+,0)+\eta(-,0)}} e^{i z \cdot V_+(\eta)} 
+  \sqrt{\frac{\eta(-,1)}{\eta(+,0)+\eta(-,0)}}  e^{i z \cdot V_+(\eta^f)} \right) 
\right] \ .
\end{align}
We will now trade the integrations over $Q_{0,n}$ and $Q_{1,n}$ for integrals over $\mu_n$, thanks to the change of densities expressed in (\ref{eq:symmetry_Q_01}) and (\ref{eq:definition_mu_n}). We will also write $x_{0,n} + x_{1,n}=2 x_{1,n} + Be^{-\g}$ according to (\ref{eq:recurs_x1}), and write $2 x_{1,n} $ as an integral over $\mu_n$ following (\ref{eq_x1n_as_mu}). This yields
\begin{align}
\hmu_{n+1}(z)&=\exp\left[-\g - B \, e^{-\g} -\frac{\he^2}{2}  - \frac{\he^2}{2}  (\za+\zb)^2 
+ \int\dd \mu_n(\eta) \sqrt{\frac{\eta(+,0)+\eta(-,0)}{\eta(-,1)}}  \, e^{iz \cdot V_-(\eta)} 
\right. \nonumber
\\  & \left. \qquad \qquad
+ \int \dd \mu_n(\eta)\left(  
\frac{\eta(+,0)}{\sqrt{\eta(-,1)(\eta(+,0)+\eta(-,0))}}  \, e^{i z \cdot V_+(\eta)} 
+ \frac{\eta(+,0)}{\sqrt{\eta(+,1)(\eta(+,0)+\eta(-,0))}} \,  e^{i z \cdot V_+(\eta^f)} \right) 
\right. \nonumber
\\  & \left. \qquad \qquad
- \int \dd \mu_n(\eta)  \frac{\eta(+,1)+\eta(-,1)}{\sqrt{\eta(+,1)\eta(-,1)}}
 \right] \ .
\end{align}
Using the invariance under spin-flip of $\mu_n$ one can regroup the two terms in the second line of this equation; simplifying the prefactors one obtains
\begin{align}
\hmu_{n+1}(z)&=\exp\left[-\g - B \, e^{-\g} -\frac{\he^2}{2}  - \frac{\he^2}{2}  (\za+\zb)^2 
\right. \nonumber
\\  & \left. \qquad \qquad
+ \int \dd \mu_n(\eta)\left(  
\sqrt{\frac{\eta(+,0)+\eta(-,0)}{\eta(-,1)}} \left( e^{i z \cdot V_+(\eta)} + e^{i z \cdot V_-(\eta)} \right) 
-\frac{\eta(+,1)+\eta(-,1)}{\sqrt{\eta(+,1)\eta(-,1)}} \right)
 \right] \ .
\end{align}
This is a recursion equation for the reweighted measure $\mu_n$ (and its characteristic function $\hmu_n$). It will be more convenient in the following to work with the pair $(\nu_n,m_n)$; the mass $m_n$ of $\mu_n$ can be expressed as $\hmu_n(0)$, we thus obtain
\begin{align}
\label{eq:recurs_mass}
m_{n+1}&=\exp\left[-\g-B \, e^{-\g} -\frac{\he^2}{2} + m_n\int \dd\nu_n(\eta) M(\eta) \right] \ , \\
\label{eq:recurs_hnu}
\hnu_{n+1}(z)&=\exp\left[- \frac{\he^2}{2}  (\za+\zb)^2  + m_n \int \dd\nu_n(\eta) L(\eta) \left( e^{i z \cdot V_+(\eta)} + e^{i z \cdot V_-(\eta)} -2 \right) \right] \ , 
\end{align}
where we introduced the functions
\begin{align}
M(\eta) &= \frac{\sqrt{\eta(+,0)+\eta(-,0)} (\sqrt{\eta(+,1)} + \sqrt{\eta(-,1)}) -\eta(+,1)-\eta(-,1)}{\sqrt{\eta(+,1)\eta(-,1)}} \ , \\
L(\eta) &= \sqrt{\frac{\eta(+,0)+\eta(-,0)}{\eta(-,1)}}  \ .
\end{align}
The initial condition for the recursion on $(\nu_n,m_n)$ is obtained from the one on $Q_{1,1}$ given in (\ref{eq:initialcondition_Q_z}):
\begin{align}
m_1 &= \exp\left[-\g+e^{-\g}\left(\frac{2}{\sqrt{b}}-\frac{1}{b}-1\right) -\frac{\he^2}{2} \right] \ ,
\label{eq_m1}
\\
\nu_1(\eta)&= 
\sum_{q,r=0}^{\infty} \Po\left(q;\frac{e^{-\g}}{\sqrt{b}}\right) \Po\left(r;\frac{e^{-\g}}{\sqrt{b}}\right) 
\int \frac{\dd \alpha}{\sqrt{2 \pi \he^2}} e^{-\frac{\alpha^2}{\he^2}} \delta\left(\tu - V_0\left(\alpha+\frac{q-r}{2} \ln b\right) \right) \ .
\label{eq_nu1}
\end{align}

For completeness we give here the expressions of the functions we introduced in terms of the $u$-parametrization:
\begin{align}
L(\eta) &= \sqrt{1 + B \frac{\ua \uc}{\ub} + (\ua)^2 + B \ua \ub \uc  }\ , \\
M(\eta) &= \sqrt{1 + B \frac{\ub \uc}{\ua} + \frac{1}{(\ua)^2} + B \frac{\uc}{\ua \ub} } - \frac{1}{\ua}  \\ 
& + \sqrt{1 + B \frac{\ua \uc}{\ub} + (\ua)^2 + B \ua \ub \uc } - \ua \ , \\
C(\eta) &= 2 \frac{\left(1 + B \frac{\ub \uc}{\ua} \right) \left(1 + B \frac{\ua \uc}{\ub} \right)}{\ua + B \ub \uc + \frac{1}{\ua} + B \frac{\uc}{\ub}}\ .
\end{align}

\subsubsection{A Gaussian approximation for the quasi-hard fields}
\label{sec_Gaussian}

We have obtained above the recursion equations (\ref{eq:recurs_mass},\ref{eq:recurs_hnu}) for the scalar $m_n$ and the probability distribution $\nu_n$, complemented by the initial conditions (\ref{eq_m1},\ref{eq_nu1}). We will now discuss the possibility to solve numerically this recursion with a population representation of $\nu_n$, and its advantages with respect to the direct resolution in terms of $Q_{w,n}$. To do so let us first rewrite the recursion equation (\ref{eq:recurs_hnu}) on $\nu_n$ as
\beq
\hnu_{n+1}(z)=\exp\left[- \frac{\he^2}{2}  (\za+\zb)^2\right] \exp\left[\int\dd \pi_n(\tu)(e^{iz \cdot \tu}-1)\right] \ ,
\label{eq:recurs_hnu2}
\eeq
where we have introduced a measure $\pi_n$ of total mass we shall denote $\lambda_n$, according to
\begin{align}
\pi_n(\tu)&=m_n\int\dd\nu_n(\eta) L(\eta) \left(\delta(\tu-V_+(\eta))+ \delta(\tu-V_-(\eta))\right) \ , \\
\lambda_n &=\int\dd\pi_n(\tu) = 2m_n\int \dd\nu_n(\eta) L(\eta) = m_n\int \dd\nu_n(\eta) \sqrt{\frac{\eta(+,0) + \eta(-,0)}{\eta(+,1) \eta(-,1)}} (\sqrt{\eta(+,1)} + \sqrt{\eta(-,1)} ) \ ,
\label{eq_lambda_n}
\end{align}
where the last expression of $\lambda_n$ has been obtained by symmetrizing the integrand.

According to the equation (\ref{eq:recurs_hnu2}) a random variable $\tu$ drawn from $\nu_{n+1}$ can be decomposed as the sum of two random variables, one Gaussian distributed and the other with a compound Poisson distribution. More explicitly one has the following equality in distribution, $\tu \eqd \alpha (\ea + \eb) + \sum_{i=1}^p \tu_i$ where $\alpha$ is a Gaussian with zero mean and variance $\he^2$, $p$ is extracted from a Poisson law of mean $\lambda_n$, and the $\tu_i$'s are i.i.d. copies extracted from $\pi_n /\lambda_n$. If $\nu_n$ is known as an empirical distribution over a sample then it is possible to draw $\tu_i$ from the probability law $\pi_n/\lambda_n$ by extracting a field $\eta$ in the population representing $\nu_n$ with a probability proportional to $L(\eta)$, and then setting $\tu_i=V_\delta(\eta)$ with $\delta = \pm$ with equal probability $1/2$. It seems then possible to use this distributional interpretation to solve numerically the recursion on $(m_n,\nu_n)$. However this is doable in practice only if $\lambda_n$ remains bounded when $n$ grows, otherwise one falls back on the problem we wanted to avoid of having to manipulate a diverging number of summands. As a matter of fact the reweighting has not offered a free lunch from this point of view: it turns out that $\lambda_n$ diverges if and only if $x_{1,n}$ does, in other words if and only if $\g < \g_{\rm r}(b)$. This statement is a consequence of the bounds $c_-(b) x_{1,n} \le \lambda_n \le c_+(b) x_{1,n}$, where $c_\pm(b)$ are positive constants, the proof of which we defer to the Appendix~\ref{app_inequalities} for the sake of readability.

Fortunately the reweighting procedure we followed will help us to handle the divergence of $\lambda_n$ more easily than the one of $x_{1,n}$ in the direct recursion. Indeed the divergence of $\lambda_n$ comes from the contributions of fields for which $L(\eta)$ becomes very large; the crucial point is that these $\eta$ yield very small values of $V_\pm(\eta)$, we can thus make a Gaussian approximation for this sum of a very large number of very small random variables. To put this idea at work we rewrite (\ref{eq:recurs_hnu2}) by decomposing it as
\beq
\hnu_{n+1}(z)=\exp\left[- \frac{\he^2}{2}  (\za+\zb)^2\right] \hnu^{(\leq)}_{n+1}(z)\, \hnu^{(>)}_{n+1}(z) \ ,
\label{eq_decomposition_hnu}
\eeq
with
\begin{align}
\hnu^{(\leq)}_{n+1}(z) &=\exp\left[m_n\int \dd \nu_n(\eta) L(\eta)\left(e^{i z \cdot V_+(\eta)} + e^{i z \cdot V_-(\eta)} -2\right)\ind\left[L(\eta) \leq\xi_n\right]\right] \ , \\
\hnu^{(>)}_{n+1}(z)&=\exp\left[m_n\int \dd \nu_n(\eta) L(\eta) \left(e^{i z \cdot V_+(\eta)} + e^{i z \cdot V_-(\eta)} -2\right)\ind\left[L(\eta)>\xi_n\right]\right] \ ,
\end{align}
where $\xi_n$ is a threshold that is arbitrary for the moment, we shall specify it later on. The decomposition (\ref{eq_decomposition_hnu}) means that under the law $\nu_{n+1}$ the random variable $\tu$ is the sum of the Gaussian random variable described previously and of two random variables, one with the law $\nu_{n+1}^{(\leq)}$, the other with the law $\nu_{n+1}^{(>)}$.

We describe the distribution $\nu^{(\leq)}_{n+1}$ using the interpretation explained above, defining
\begin{align}
\pi_n^{(\leq)}(\tu)&=m_n\int\dd\nu_n(\eta) L(\eta) \left(\delta(\tu-V_+(\eta))+ \delta(\tu-V_-(\eta))\right)\ind\left[L(\eta)\leq\xi_n\right] \ ,\\
\lambda_n^{(\leq)} &=\int\dd\pi_n^{(\leq)}(\tu) = 2m_n\int \dd\nu_n(\eta) L(\eta) \ind\left[L(\eta) \leq\xi_n\right] \ .
\label{eq:def_lambdaleq}
\end{align}
Under the law $\nu_{n+1}^{(\leq)}$ the variable $\tu$ obeys the distributional equality $\tu=\sum_{i=1}^p \tu_i$ where $p$ is a Poisson variable of mean $\lambda_n^{(\leq)}$, and the $\tu_i$'s are i.i.d copies extracted from $\pi_n^{(\leq)}/\lambda_n^{(\leq)}$.

The contribution $\nu^{(>)}_{n+1}$ is instead approximated by a multivariate Gaussian $\G(\Vbar_n,\Sigma_n)$ with $\Vbar_n$ and $\Sigma_n$ the mean and the covariance matrix of $\nu_{n+1}^{(>)}$, computed by taking derivatives of $\ln \hnu^{(>)}_{n+1}$ with respect to $z$:
\begin{align}
    \label{eq:recurs_Mn}
    &\Vbar_n = m_n\int\dd\nu_n(\eta)L(\eta) \ind\left[L(\eta) > \xi_n\right](V_+(\eta) + V_-(\eta)) \ ,\\
    \label{eq:recurs_Sigman}
    &\Sigma_n^{(a),(b)} = m_n\int\dd \nu_n(\eta)L(\eta) \ind\left[L(\eta) > \xi_n\right]
(V_+(\eta)^{(a)}V_+(\eta)^{(b)} + V_-(\eta)^{(a)}V_-(\eta)^{(b)} ) \ ,
\end{align}
for $a,b \in \{1,2,3\}$. As $V_-(\eta)=(V_+(\eta))^f$ several components of $\Vbar_n$ and $\Sigma_n$ vanish, namely $\Vbar_n^{(1)}=\Vbar_n^{(2)}=\Sigma_n^{(1),(3)}=\Sigma_n^{(2),(3)}=0$. 

Replacing $\nu^{(>)}_{n+1}$ by a Gaussian is an approximation, that amounts to neglect the cumulants of order larger than 2, the accuracy of which is controlled by the cutoff $\xi_n$. The larger is $\xi_n$ the better the truncation is, because a smaller part of the full law $\nu_{n+1}$ is treated approximatively, but the price to pay is a simultaneous increase of $\lambda_n^{(\leq)}$, the average number of fields that must be summed in the description of $\nu_{n+1}^{(\leq)}$. A compromise needs thus to be found between these two effects, we explain below how we fixed $\xi_n$ in practice.

\subsubsection{Algorithmic implementation}

We now give an explicit description of the algorithm we implemented to solve the recursion equations (\ref{eq:recurs_mass},\ref{eq:recurs_hnu}) for $m_n$ and $\nu_n$. Suppose that at the $n$-th step of the iteration we have an estimation of $m_n$ and of $\nu_n$, with $\nu_n$ represented as a population of fields:
\beq
\nu_n(\eta)\simeq\frac{1}{\N}\sum_{i=1}^{\N}\delta(\eta-\eta_i) \ .
\label{eq_nu_population}
\eeq
One can evaluate the average of an arbitrary function $A$ with respect to $\nu_n$ as
\beq
\int \dd \nu_n(\eta) A(\eta)\simeq \frac{1}{\N} \sum_{i=1}^{\N}A(\eta_i) \ ,
\eeq
and in particular compute in this way $m_{n+1}$ from (\ref{eq:recurs_mass}). We further assume that the fields $\eta_i$ have been sorted by increasing values of $L(\eta)$, and translate the cutoff $\xi_n$ by defining the index $\N_n$ such that $L(\eta_{\N_n}) \leq \xi_n < L(\eta_{\N_n +1 })$. The integrals where the indicator function $\ind[L(\eta) \le \xi_n ]$ (resp. $\ind[L(\eta) > \xi_n ]$) can thus be translated as sums over the population elements from $1$ to $\N_n$ (resp. from $\N_n +1$ to $\N$), which allows to compute easily $\lambda_n^{(\leq)}$ from (\ref{eq:def_lambdaleq}), and $\Vbar_n$ and $\Sigma_n$ from (\ref{eq:recurs_Mn}) and (\ref{eq:recurs_Sigman}).

Each of the $\N$ elements of the new population representing $\nu_{n+1}$ is then generated independently of the others, by translating equation (\ref{eq_decomposition_hnu}) as follows:
\begin{itemize}
\item draw an integer $p$ from the Poisson law of mean $\lambda_n^{(\leq)}$.
\item extract $i_1,\dots, i_p$ independently in $\{1,\dots,\N_n\}$ with probability proportional to $L(\eta_i)$ (this can be done efficiently by precomputing a cumulative table).
\item insert in the new population $\tu=V_{\delta_1}(\eta_{i_1})+\dots+V_{\delta_p}(\eta_{i_p}) + \alpha (\ea + \eb) + \Vbar_n + g$ where the $\delta_i$'s are $\pm$ with equal probability, $\alpha$ is a centered Gaussian random variable of variance $\he^2$, and $g$ is a centered three-dimensional Gaussian vector with covariance matrix $\Sigma_n$.
\end{itemize}
We can then compute the reduced overlap $\tC_{n+1}$ from (\ref{eq:tC_computedfrom_nu}), and sort the elements of the new population according to their values of $L(\eta)$. 

In practice we chose the threshold $\xi_n$ (or equivalently $\N_n$) in an adaptive way: for each iteration step we took the largest $\N_n \le \N$ that gave $\lambda_n^{(\leq)} \le \lambda$, where $\lambda$ is a parameter fixed beforehand. The accuracy of this numerical procedure is thus controlled by $\N$, the approximation in (\ref{eq_nu_population}) being better and better as $\N$ grows, and by $\lambda$, the Gaussian truncation being more precise when $\lambda$ is larger. Obviously the memory and time requirements of the procedure also increase with $\N$ and $\lambda$; the numerical results presented below have been obtained with population sizes $\N$ between $10^6$ and $10^7$, and $\lambda$ around 20, we checked that the conclusions were not modified, within our numerical accuracy, by modifying these values in a reasonable range.

\subsection{Results}
\label{sec_largek_results}

\begin{figure}
\begin{center}
\includegraphics[width=7.7cm]{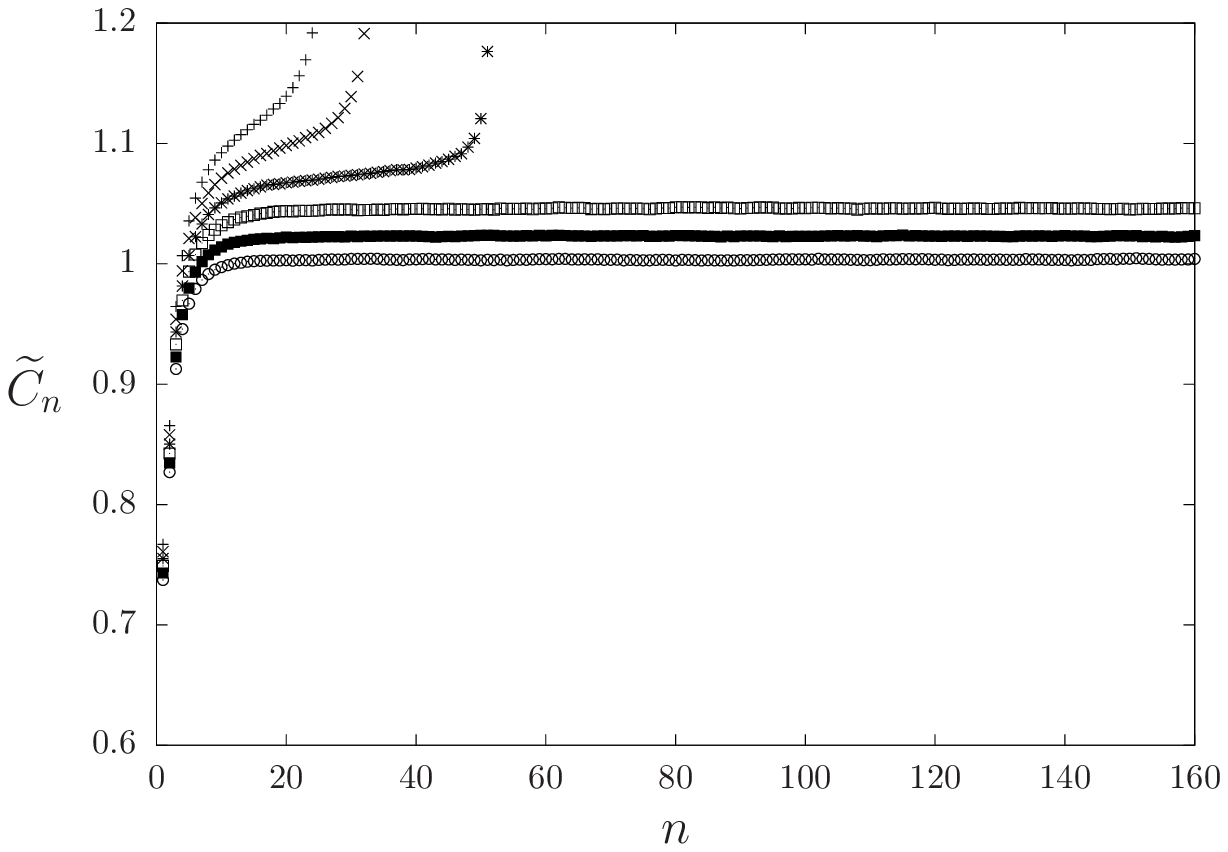}
\hspace{1cm}
\includegraphics[width=8cm]{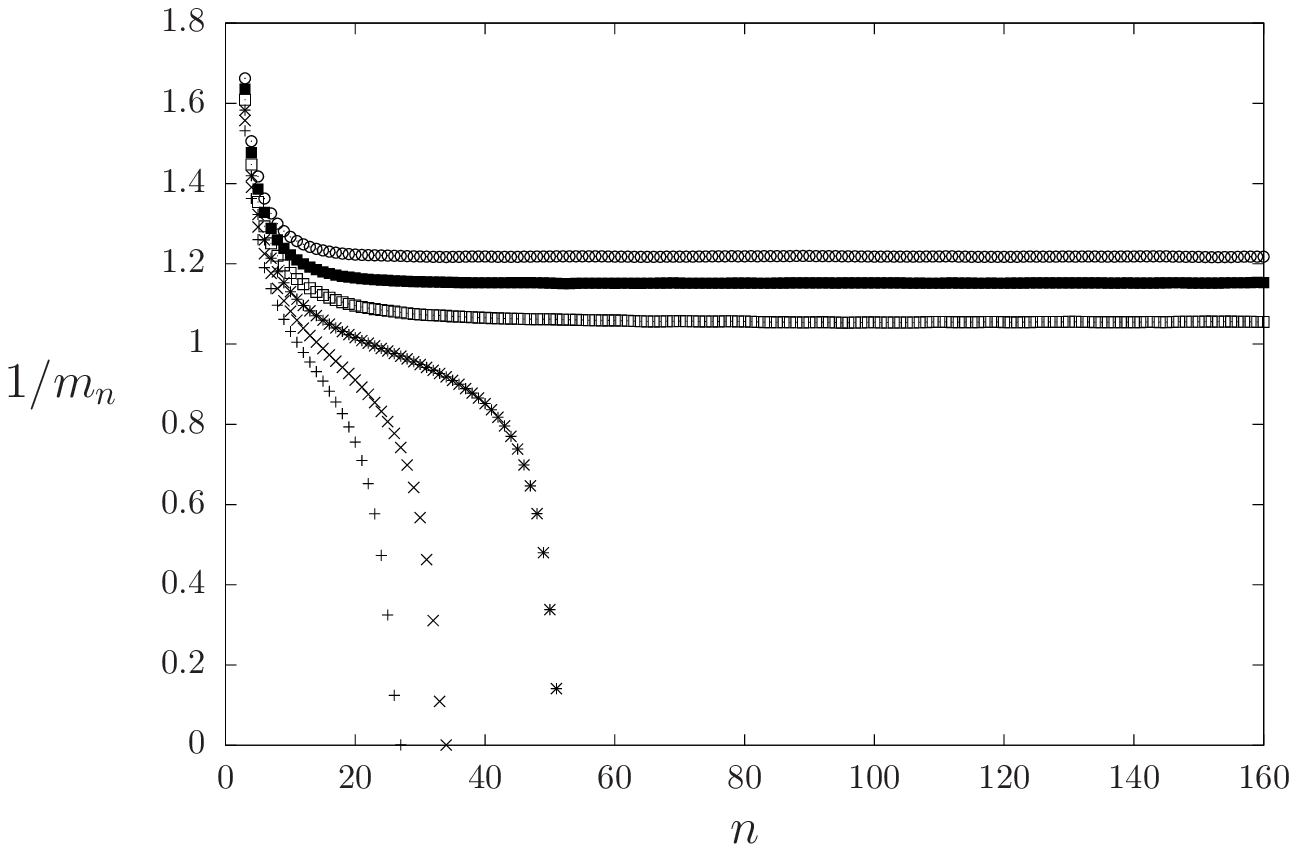}
\caption{The reduced correlation function $\tC_n$ (left panel) and the inverse of the mass $m_n$ of $\mu_n$ (right panel) for $b=0.4$, $\he=0$, and from left to right in both panels, $\g=0.95$, $0.96$, $0.97$, $0.98$, $0.99$, $1$.}
\label{fig:masstCvst_b0p4}
\end{center}
\end{figure}

In Fig.~\ref{fig:masstCvst_b0p4} we complete our study of the case $b=0.4$, $\he=0$ that was started in Figs.~\ref{fig:ath_tC} and \ref{fig:xtCvst_b0p4_g1p35}. The reweighting procedure allows now to investigate the regime $\g < \g_{\rm r} \approx 1.378$; as displayed on the left panel of Fig.~\ref{fig:masstCvst_b0p4} the large distance limit $\tC$ remains bounded for values of $\g$ down to $0.98$, these results being reported as a function of $\g$ in the right panel of Fig.~\ref{fig:ath_tC}. Further simulations allowed us to pinpoint more precisely $\g_{\rm d}(b=0.4,\he=0) \approx 0.977$, as the largest value of $\g$ for which $\tC_n$ diverges. As shown in the right panel of Fig.~\ref{fig:masstCvst_b0p4} the condition of divergence of $\tC_n$ coincides, within our numerical accuracy, with the divergence of $m_n$. The value of the large $n$ limit of $m_n$ is seen to be close to 1 when $\g$ reaches $\g_{\rm d}$ from above (see the plateau in the right panel of Fig.~\ref{fig:masstCvst_b0p4}), an observation that we also made for the other values of $b$ we investigated. We have given analytical arguments in~\cite{BuSe19} that indeed the plateau value of $m_n$ is exactly equal to 1 at $\g_{\rm d}$ for the uniform measure, our numerical results suggest that this remains true when $(b,\he) \neq (1,0)$, even if we do not have analytical support for this assumption in the general case.

\begin{figure}
\begin{center}
\includegraphics[width=8cm]{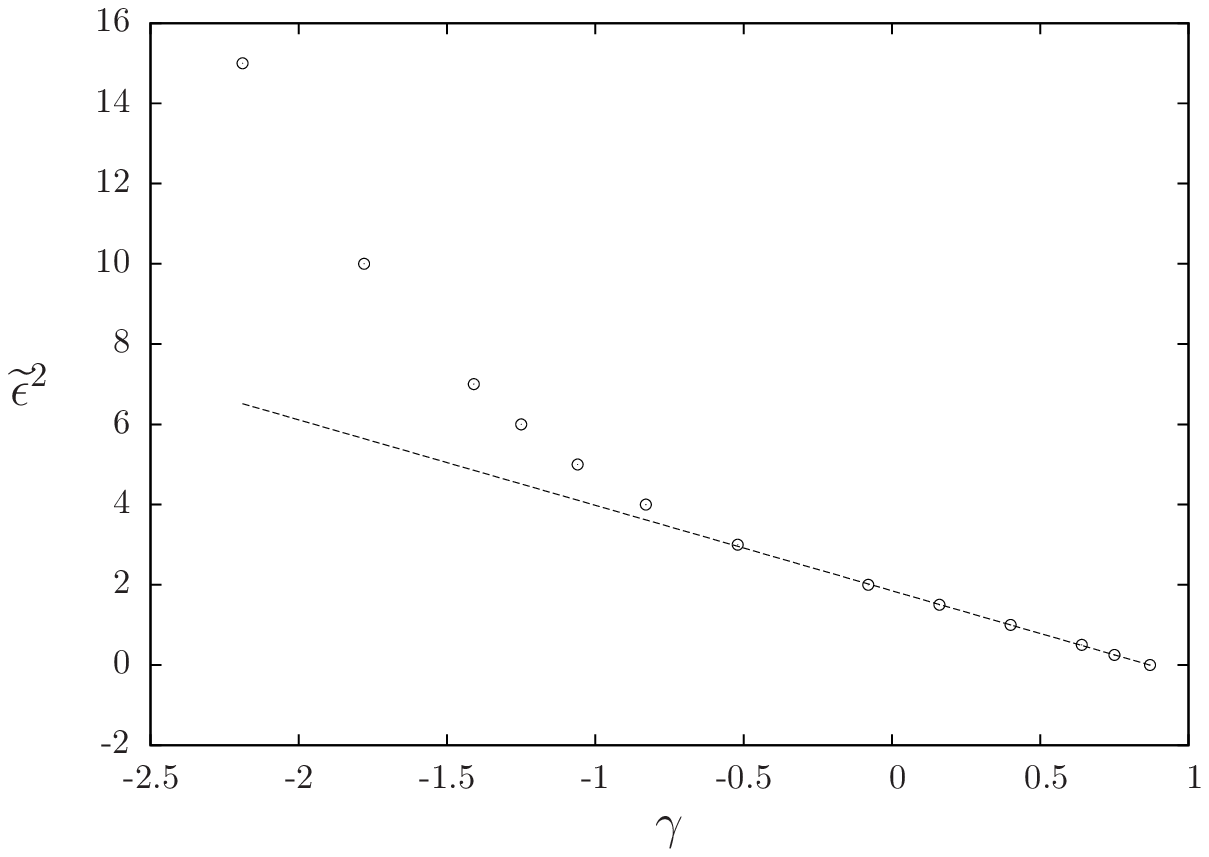}
\hspace{1cm}
\includegraphics[width=8cm]{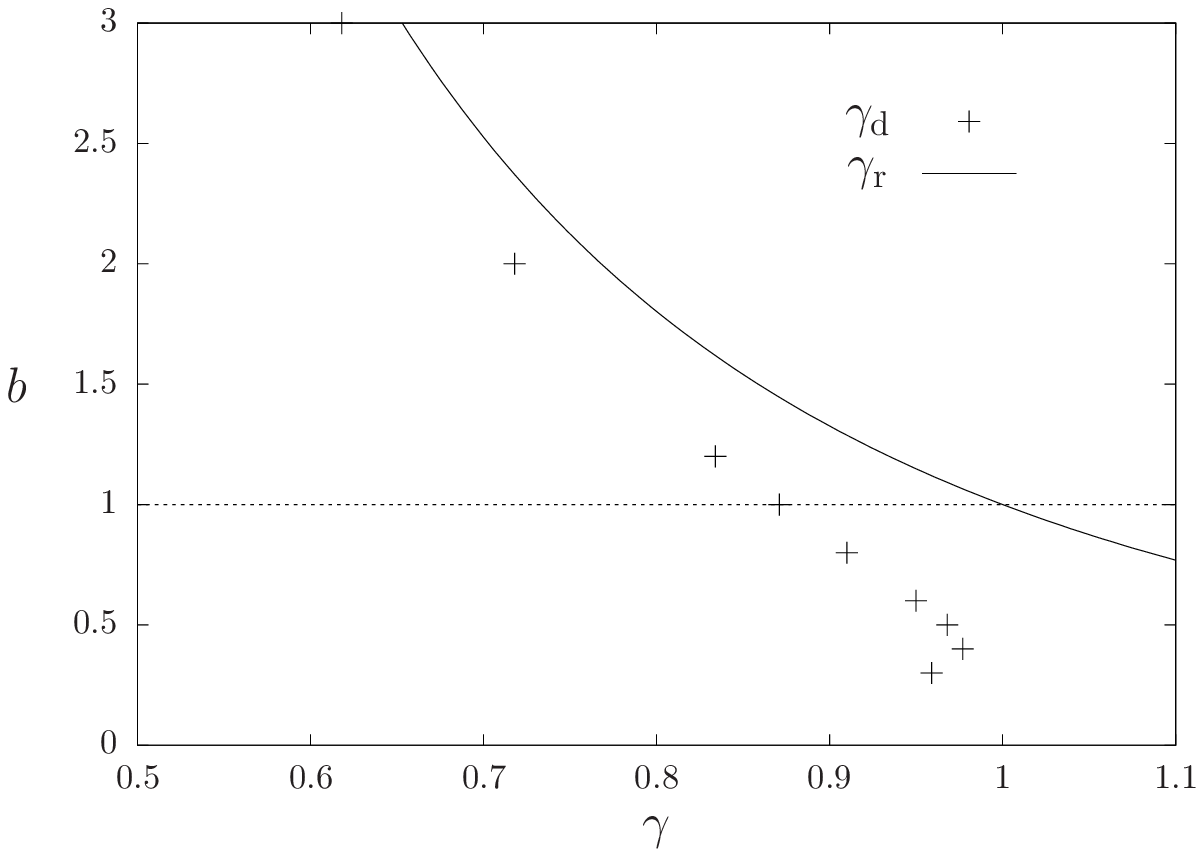}
\caption{Left panel: $\g_{\rm d}$ as a function of $\he^2$ for $b=1$, the line being a guide to the eye.
Right panel: $\g_{\rm d}$ as a function of $b$ for $\he=0$, the solid line corresponding to the rigidity upperbound $\g_{\rm r}(b)$.}
\label{fig_pd_eps}
\end{center}
\end{figure}

\begin{figure}
\begin{center}
\includegraphics[width=7cm]{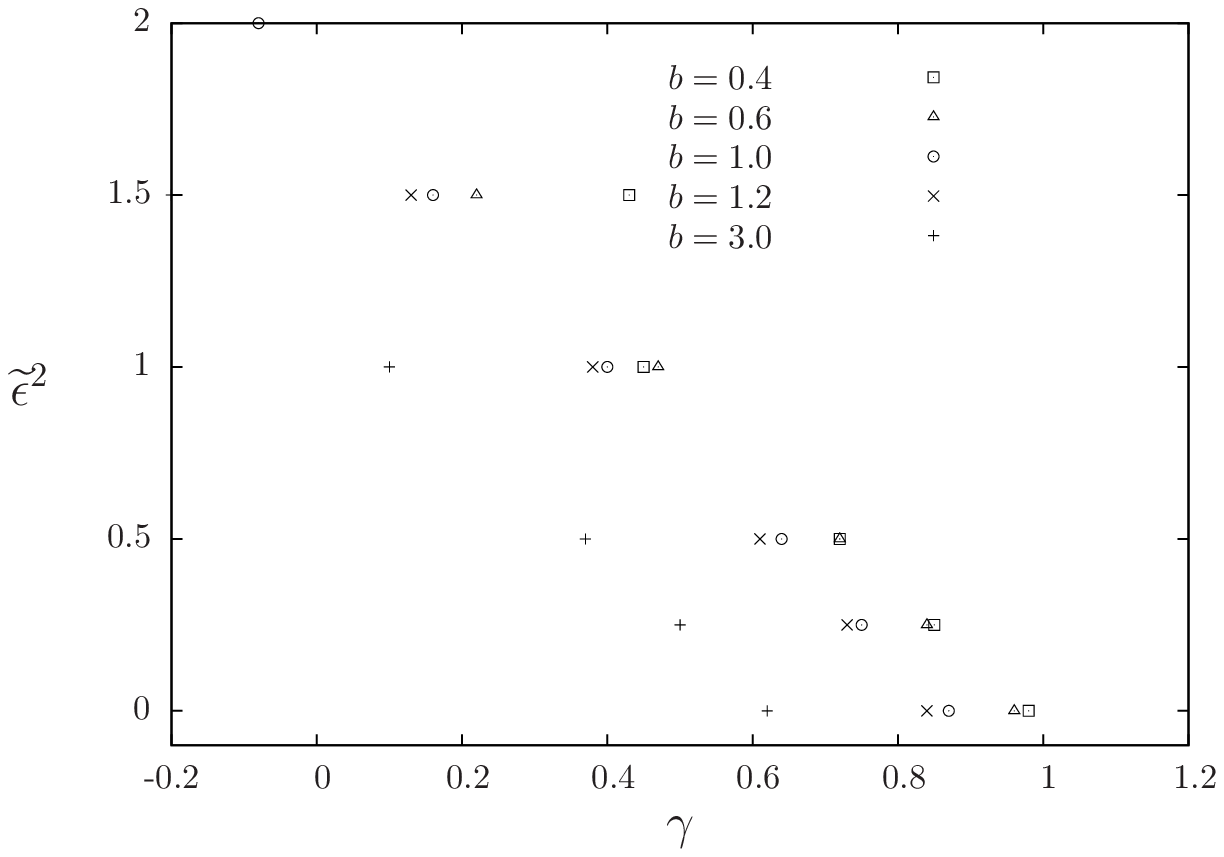}
\caption{$\g_{\rm d}$ as a function of $\he$ for various $b$. }
\label{fig:bvsgamma}
\end{center}
\end{figure}

We have repeated this procedure of determination of $\g_{\rm d}(b,\he)$ for various values of $b$ and $\he$, and we present now the phase diagrams obtained in this way. Consider first the left panel of Fig.~\ref{fig_pd_eps}, which deals with the case $b=1$, i.e. the bias factorized over the hyperedges considered in~\cite{BuRiSe19}. We see that for all values of $\he \neq 0$ one has $\g_{\rm d}(1,\he) < \g_{\rm d}(1,0) =\g_{\rm d,u}$, i.e. this bias has, in the large $k$ limit, a detrimental effect on the dynamic phase transition that is pushed to lower values with respect to the one of the uniform measure. In the right panel of Fig.~\ref{fig_pd_eps} we have plotted instead the threshold $\g_{\rm d}$ as a function of $b$ for $\he=0$; one sees now that decreasing $b$ below 1 (that corresponds to the uniform measure and is marked as an horizontal dashed line on the figure) has a beneficial effect with an increase of $\g_{\rm d}$. The largest value we could reach was for $b=0.4$, decreasing $b$ further below reduces again $\g_{\rm d}$. The lowest value of $b$ we could investigate was $b=0.3$, for $b<0.3$ we encountered numerical accuracy problems, the distribution $\nu_n$ exhibiting strong fluctuations that prevented an accurate representation as a population. Finally in Fig.~\ref{fig:bvsgamma} we have checked that the parameter $\he$ has a detrimental effect also for values of $b \neq 1$, we found indeed that $\g_{\rm d}(b,\he) < \g_{\rm d}(b,0)$ when $\he \neq 0$, for all the values of $b$ we considered. This leads us to the conclusion that, within the biasing strategy we considered in the large $k$ limit, the optimal choice of parameters is $\he =0$ and $b \approx 0.4$, yielding a constant $\g_{\rm d} \approx 0.977$, strictly larger than the one of the uniform case, $\g_{\rm d,u} \approx 0.871$. The observation of the right panel of Fig.~\ref{fig_pd_eps} allows to justify the choice made in the beginning of the section for the scaling of $b$. Since we obtain an optimum for the dynamical threshold when $b$ is finite, we expect to have a smaller dynamical threshold if we choose a different scaling (i.e $b$ going to $0$ or to $+\infty$ in the large $k$ limit).

\section{Conclusions}
\label{sec_conclusion}

We have performed in this article a quantitative study of the biasing strategy for random constraint satisfaction problems, focusing on the $k$-hypergraph bicoloring problem and a bias coupling variables at distance 1 on the hypergraph. We have determined the dynamic transition both for finite $k$ via a numerical resolution of the 1RSB equations, and in the large $k$ limit through a partly analytic asymptotic expansion. We have shown that the increased range of soft interactions with respect to the ones factorized over the hyperedges enhance the efficiency of the bias by further pushing the dynamic transition to higher constraint densities, and in particular in the large $k$ limit we have achieved a constant $\g_{\rm d} \approx 0.977$ in the asymptotic expansion $\alpha = 2^{k-1} (\ln k + \ln \ln k + \g)$, strictly greater than the one obtained in the absence of bias, $\g_{\rm d,u} \approx 0.871$. Let us sketch now some possible directions for future research that these results suggest. 

The improvement in the asymptotic behavior of the dynamic transition is certainly moderate, as it occurs in the third order of the asymptotic expansion, the constant $\g_{\rm d}$ remaining itself smaller than the rigidity threshold $\g_{\rm r,u}=1$ of the uniform measure. However, the conceptual link between this transition and the important algorithmic gap explained in the introduction justifies for us the efforts devoted to achieve this improvement, and calls for further investigations. The natural question that arises is to determine the optimal asymptotic scaling of $\alpha_{\rm d}$ that can be achieved for a biased measure incorporating soft interactions of an arbitrary but finite range, and in particular at which order of the asymptotic expansion does $\alpha_{\rm d}$ exceeds $\alpha_{\rm d,u}$. This seems quite a challenge in this fully general form, but partial results could certainly be obtained by extending our study, for instance considering more general forms of $\psi(p)$ than the one of equation (\ref{eq_bias_beps}), retaining more information on the local configuration than just the number of forcing clauses around one variable, or increasing the range of interactions to distance 2. Both positive or negative results, i.e. the impossibility to increase $\alpha_{\rm d}$ beyond the third order term of the asymptotic expansion, would shed light on the intrisic difficulty of random CSPs, putting barriers for larger and larger families of algorithms, in the spirit of~\cite{GaSu14,CoHaHe17,Hetterich}. Finally it would also be interesting to extend this study to other CSPs, in particular the $k$-satisfiability and the $q$-coloring problems.

\acknowledgments

We thank Federico Ricci-Tersenghi and Lenka Zdeborova for useful discussions. GS is part of the PAIL grant of the French Agence Nationale de la Recherche, ANR-17-CE23-0023-01.

\appendix

\section{Existence and uniqueness of the RS solution}
\label{app_uniqueness_RS}

In this appendix we shall show that the translationally invariant RS equation (\ref{BP_invariance_spin_transl_with_yhy}) admits a unique solution for all choices of the bias function $\psi$ that is strictly positive, $\psi(p)>0$ $\forall p\in\{0,\dots l+1\}$. 

We first remark that in the uniform case, where $\psi(p)$ is a positive constant independent of $p$, the equation (\ref{BP_invariance_spin_transl_with_yhy}) obviously admits a unique solution (with $y=1$, $\hy=2^{k-1}-2$). We will now show that the number of solutions cannot change when $\psi$ varies in its allowed domain. To achieve this we first rewrite  (\ref{BP_invariance_spin_transl_with_yhy}) in the equivalent form
\beq
G(x;\psi)=\sum_{p=0}^{l+1}\psi(p) X_p(x)=0 \ ,
\label{eq_G}
\eeq
where for simplicity we denoted $x=1/\hy$ and where the coefficients $X_p(x)$ are:
\begin{align}
    X_p(x)&=x^p\left[(2^{k-1}-k-1)\binom{l}{p} x+(k-1)\binom{l}{p-1} - \binom{l}{p} \right] \\
    &=\frac{x^p}{l+1}\binom{l+1}{p}\left[(2^{k-1}-k-1)(l+1-p)x-(l+1-kp)\right] \ ; \label{eq_X}
\end{align}
in the first line we used the convention $\binom{l}{l+1}=\binom{l}{-1}=0$.

The function $G(x;\psi)$ introduced in (\ref{eq_G}) depends smoothly on its two arguments (polynomially in $x$, and linearly in $\psi$), the number of solutions $x(\psi)$ of the equation $G=0$ can thus only change at a bifurcation point, i.e. a pair $(x;\psi)$ such that $G(x;\psi)=\partial_xG(x;\psi)=0$, otherwise the implicit function theorem allows to smoothly continue any branch of solution. As we remarked above the solution is unique when $\psi$ is independent of $p$, the uniqueness for all $\psi$ will then follow if we show the absence of solution to the bifurcation equation:
\begin{align}
\label{eq:bifurc_G}
\sum_{p=0}^{l+1}\psi(p)X_p(x)=0 \ ,\quad
\sum_{p=0}^{l+1}\psi(p)Y_p(x)=0 \ , \quad \psi(p) \ge 0 \ \ \ \forall p \ ,
\end{align}
where
\beq
Y_p(x) = \frac{\partial X_p}{\partial x} = \frac{x^{p-1}}{l+1}\binom{l+1}{p}\left[(2^{k-1}-k-1)(l+1-p)(p+1)x-(l+1-kp)p\right] \ . \label{eq_Y}
\eeq
This equation being invariant under the multiplication of $\psi$ by a positive constant we can further assume the normalization condition
\beq
\label{eq:bifurc_G_normalization}
\sum_{p=0}^{l+1}\psi(p)=1 \ .
\eeq
For a given value of $x$, the existence of a $\psi$ satisfying (\ref{eq:bifurc_G},\ref{eq:bifurc_G_normalization}) is equivalent to the origin of $\mathbb{R}^2$ being in the convex hull of the $l+2$ points of coordinates $\begin{pmatrix} X_p(x) \\ Y_p(x) \end{pmatrix}$ for $p=0,\dots,l+1$. We can then invoke the Caratheodory theorem that states that any point of the convex hull of a set $A \subset \mathbb{R}^d$ can be written as the convex combination of $d + 1$ points of $A$. Here $d=2$, so the absence of solutions of (\ref{eq:bifurc_G},\ref{eq:bifurc_G_normalization}) follows from the impossibility to satisfy, for any $p,q,r\in\{0,\dots, l+1\}$, the system
\begin{align}
& \alpha_p X_p(x) + \alpha_q X_q(x) + \alpha_r X_r(x)  = 0 \ , \\
& \alpha_p Y_p(x) + \alpha_q Y_q(x) + \alpha_r Y_r(x)  = 0 \ , \\
& \alpha_p \ge 0 , \quad \alpha_q \ge 0 , \quad \alpha_r \ge 0 , \quad \alpha_p+\alpha_q+\alpha_r > 0 \ .
\end{align}
This is equivalent to the three quantities $X_p(x)Y_q(x) - X_q(x)Y_p(x)$, $X_q(x)Y_r(x) - X_r(x)Y_q(x)$ and $X_r(x)Y_p(x) - X_p(x)Y_r(x)$ being of the same sign; using the expressions (\ref{eq_X},\ref{eq_Y}) of $X$ and $Y$ we have checked the impossibility of this condition, for all $x>0$ and all triplets $p,q,r$, which concludes the reasoning.

\section{An inequality}
\label{app_inequalities}

We provide in this Appendix a proof of the bounds $c_-(b) x_{1,n} \le \lambda_n \le c_+(b) x_{1,n}$ that we used in Sec.~\ref{sec_Gaussian}. We start by stating some inequalities that are fulfilled by the messages $\eta$ in the support of $\nu_n$, and that are consequences of the BP equation (\ref{eq_f_b_eps0}). They are more compactly stated in terms of the $u$-parametrization; from (\ref{eq_BP_u}) one obtains indeed
\beq
\frac{\ub \uc}{\ua} = \prod_i \frac{\heta_i(+,0)}{\heta_i(+,0) + \heta_i(+,1)} \ , \qquad
\frac{\ua \uc}{\ub} = \prod_i \frac{\heta_i(-,0)}{\heta_i(-,0) + \heta_i(-,1)} \ ,
\eeq
which allows to conclude that $\frac{\ub \uc}{\ua} \le 1$ and $\frac{\ua \uc}{\ub} \le 1$, for all the $\eta$'s in the support of $\nu_n$.

Consider now the expressions (\ref{eq_x1n_as_mu}) for $x_{1,n}$ and (\ref{eq_lambda_n}) for $\lambda_n$; the ratio of the integrands in these two equations reads
\beq
2 \sqrt{\eta(+,0)+\eta(-,0)} \frac{\sqrt{\eta(+,1)} + \sqrt{\eta(-,1)}}{\eta(+,1)+\eta(-,1)} = 
\sqrt{\frac{\eta(+,0)+\eta(-,0)}{\eta(+,1)+\eta(-,1)}}
\left(2 \sqrt{\frac{\eta(+,1)}{\eta(+,1)+\eta(-,1)}} + 2 \sqrt{\frac{\eta(-,1)}{\eta(+,1)+\eta(-,1)}} \right) \ .
\nonumber
\eeq
The parenthesis in the right hand side of this equation is of the form $2(\sqrt{x} + \sqrt{1-x})$ for some $x$ in $[0,1]$, which is necessarily in the interval $[2,2\sqrt{2}]$. The prefactor in front of the parenthesis can be written, in terms of the $u$-parametrization,
\beq
\sqrt{\frac{1}{1+(\ua)^2}\left(1+ B \, \ua \ub \uc + (\ua)^2 + B \frac{\ua \uc}{\ub}  \right)} \ .
\label{eq_prefactor}
\eeq
Consider first the case $b \le 1$, i.e. $B \ge 0$; as the components of $u$ are non-negative the expression in (\ref{eq_prefactor}) is certainly lower bounded by $1$. Moreover the bounds $\frac{\ub \uc}{\ua} \le 1$ and $\frac{\ua \uc}{\ub} \le 1$ imply that it is upper bounded by $\sqrt{1+B}=1/\sqrt{b}$. The case $b \ge 1$, $B \le 0$ can be treated similarly, with now the expression in (\ref{eq_prefactor}) being in the interval $[1/\sqrt{b},1]$. Combining these observations we obtain finally $c_-(b) x_{1,n} \le \lambda_n \le c_+(b) x_{1,n}$, with for $b\le 1$ $c_-(b)=2$, $c_+(b)=2\sqrt{2/b}$, and for $b\ge 1$ $c_-(b)=2/\sqrt{b}$, $c_+(b)=2\sqrt{2}$.

\bibliography{biblio}

\end{document}